\documentclass[useAMS,usenatbib,usegraphicx]{mn2e}

\title[Frequency-dependent time delays in four
blazars]
{Frequency-dependent time delays for strong outbursts in four
blazars from the Mets\"ahovi and UMRAO monitoring databases}
\author[T. B. Pyatunina et al. ]
  {T. B. Pyatunina$^1$\footnote{Deceased in August 2005.}, N. A. Kudryavtseva$^{2,3}$, D. C. Gabuzda$^4$,
  S. G. Jorstad$^5$,
\newauthor M. F. Aller$^6$, H. D. Aller$^6$, H. Ter\"asranta$^7$\\
  $^1$Insitute of Applied Astronomy, St. Petersburg, Russia\\
  $^2$Max-Plank-Institut f\"ur Radioastronomie, Auf dem H\"ugel 63, Bonn
53121, Germany\\
  $^3$St.-Petersburg State University, Petrodvoretz, St.-Petersburg, Russia\\
  $^4$Physics Department, University College Cork, Cork, Ireland\\
  $^5$Institute for Astrophysical Research, Boston University, Boston, USA\\
  $^6$Astronomy Department, Dennison Building, University of Michigan, USA\\
  $^7$Mets\"ahovi Radio Observatory, Helsinki University of Technology, Finland}
\date{}

\def\LaTeX{L\kern-.36em\raise.3ex\hbox{a}\kern-.15em
    T\kern-.1667em\lower.7ex\hbox{E}\kern-.125emX}

\begin{document}


\maketitle

\begin{abstract}
The combined data of the University of Michigan Radio Astronomy Observatory
and Mets\"ahovi Radio Observatory provide us with radio light curves for
Active Galactic Nuclei monitored by both observatories from 4.8 to 37~GHz
covering time intervals up to $\sim$~25 years. We consider here such
composite light curves for four $\gamma$-ray blazars that have been nearly
continuously monitored at both observatories: 0458--020, 0528+134, 1730--130
and 2230+114.  We have decomposed the most prominent outbursts in the light
curves of these four blazars into individual components using
Gaussian model fitting, and estimated the epochs, amplitudes, and half-widths
of these components as functions of frequency. We attempt to distinguish
``core outbursts,'' which show frequency-dependent time delays and are
associated with brightening of the core, from ``jet outbursts,'' which appear
nearly synchronously at all frequencies and are accompanied by the emergence
of new jet components and their subsequent evolution. The outbursts in
0528+134 and 2230+114 display fine structure and consist of individual
sub-outbursts. Available 43~GHz VLBA images allow us to identify only one pure
core outburst (in 2230+114) and one pure jet outburst (0458--020). Most of the
outbursts analyzed are mixed, in the sense that they display
frequency-dependent time delays (i.e., they are optically thick) and are
associated with the eventual emergence of new jet components. The maxima of
the jet and mixed outbursts probably correspond to epochs when newly ejected
components become fully optically thin. These epochs are also marked by a
significant increase in the angular velocities of the ejected components.
There is evidence that the outbursts in 2230+114 repeat every
$8.0\pm0.3$~years, with the positions of individual sub-outbursts being
preserved from one quasi-periodic eight-year cycle to another, even though
their amplitudes vary by more than a factor of two.  Preliminary estimates
of the total durations of possible activity cycles based on an analysis of
total flux-density variations and all available VLBI data are given for the
remaining sources.
\end{abstract}


\section{Introduction}

The term ``blazar'' is sometimes used to refer to the subset of core-dominated
Active Galactic Nuclei (AGNs) that display high-amplitude variability over a
wide range of wavebands, testifying that one of the relativistic jets in these
objects is oriented close to the line of sight toward the Earth. The
variability of blazars is a very complex phenomenon, and many different
processes can take part, such as activity of the central engine, the
propagation of primary perturbations down the jet, the evolution of the jet
itself and various propagation effects (e.g. interaction with the surrounding
medium) (Marscher 1996). During the last few years, important progress
has been made in understanding many aspects of these phenomena through
theoretical simulations (see Gomez et al. 2004 and references therein),
investigations of total flux variability in the X-ray (Pian 2003), $\gamma$
(Kranich 2003) and radio (Aller et al. 1999, Ter\"asranta, Wiren \& Koivisto
2003) and various VLBI studies (e.g. Britzen et al. 1999a; Gomez et al. 2000;
Lister \& Homan 2005; Jorstad et al. 2005).

Nevertheless, our understanding of the global,
long-term development of the activity remains very fragmentary.  The detection
of periodicity in the activity of the blazars OJ287 ($P\sim12$~yr,
Pursimo et al. 2000), 0059+581 ($P\sim4$~yr, Pyatunina et al. 2003), 0202+149
($P\sim4$~yr, Pyatunina et al. 2000) and 0235+164 ($P\sim5.7$~yr, Raiteri
et al. 2001) looks promising in this regard, but all attempts to estimate
exact periods for the activity have been unconvincing, even for the best
sampled case of OJ287 (Kidger 2000), indicating that the phenomenon is
probably not strictly periodic, but instead quasi-periodic. Indeed, even if
the phenomena triggering a blazar's activity were purely periodic in nature,
the observed variability would likely be at best quasi-period, since it is
the product of multiple complex processes associated with the generation and
propagation of the radio emission. Numerical simulations of a binary black-hole
model for OJ287 (Sundelius et al. 1997) show that the triggering of the
activity itself can be a complex function of time (see below), making the
observed variability even more complex. Nevertheless, analysis of such
quasi-periodicity can be an important diagnostic tool for modelling the energy
generation in AGNs and investigating binary black-hole systems.  Below we
propose an alternative approach to investigating the global evolution of
blazars.

Physically, the activity of a source can be described using two independent
time scales: $T_{act}$, the characteristic time scale for the activity of
the central engine,
and $T_{ev}$, the time scale for evolution of the jet after it has been
disturbed by a perturbation at it's base. The time scale $T_{ev}$ includes
a whole sequence of events, from the initial appearance of the perturbation
to the time when the propagating perturbation fades and merges into the
quiescent jet emission. At present, it may be difficult to define this
important time scale observationally. The time scale $T_{act}$ can be
thought of as the duration of one complete ``activity cycle,'' which we
propose to define as the characteristic time interval between two successive
events marking onsets of phases of activity. With this definition, for
example, a blazar that showed a pattern of being active for five years,
then quiescent for ten years, then active five years, and so forth, would
have $T_{act}\simeq 15$~yrs.
If the time scale for activity exceeds the time
scale for the jet's evolution ($T_{act}\ge T_{ev}$), two subsequent activity
cycles should be easily distinguishable; otherwise, manifestations of core
activity and jet evolution associated with different cycles can be superimposed
and complicate the observed picture. It should also be borne in mind that
the observed $T_{act}$ may be affected by the source's redshift, while the
observed $T_{ev}$ may be affected by both the redshift and the Doppler factor
of the jet (for more detail, see the discussion in Lister 2001).

According to theoretical arguments (Marscher 1996; Gomez et al. 1997),
a shock induced by a primary excitation at the base of the jet is initially
manifest in radio light curves as an outburst that is delayed at lower
frequences due to the combined effects of the frequency stratification of the
emitting electrons, non-zero opacity and light-travel delays. Such time-delayed
outbursts are associated observationally on milliarcsecond (mas) scales with
brightening of the VLBI core, and can be considered ``core'' outbursts. The
propagation of the shock downstream in the jet is revealed via the emergence
of a new optically thin jet component (or components) in the VLBI image,
sometimes accompanied by ``jet'' outbursts, which evolve nearly synchronously
at all frequencies in the radio light curve. The integrated monitoring data
support the division of observed outbursts into such core and jet outbursts,
at least in some sources (0202+149, Pyatunina et al. 2000; 0420--014 Zhou et
al. 2000; 0059+581, Pyatunina et al. 2003). Thus, in practice, it is reasonable
to approximate the duration of an activity cycle as the time interval between
two successive ``core'' outbursts.
The questions of how this interval varies from source to source
and how stable it is for a particular source
may be keys for our understanding of the activity's grand design.

The majority of blazars do not seem to display a well-defined activity cycle
and associated 
time scale $T_{act}$, due in part to incompleteness of their monitoring
data over sufficiently long time scales (decades) and in part due to the
complexity or nature of the underlying physical processes giving rise to
the variability (see, e.g., Lister 2001). In addition, the clarity with which
an activity cycle in a given source is manifest may be very different at
different frequencies, especially if they are fairly widely spaced.
Combined analyses of multi-epoch VLBI observations and multi-frequency light
curves for those cases when a reasonably well-defined cycle can be identified
across a range of frequencies show that the duration of the activity cycle
can vary from source to source over a wide range:
$T_{act}\sim 4$~yr for 0059+581 (z=0.643, Pyatunina et al. 2003), $\sim 12$~yr
for 0735+178 (z=0.424, Carswell et al. 1974; Agudo et al. 2002) and
$\sim 20$~yr for 1308+326 (z=0.996, Hewitt et al. 2003; Pyatunina et al. 2004).
As has been noted above, the question of the stability of the activity
time scale, or of the quasi-periodicity of the activity, in a particular
source is more complex, not only because the relevant time scales can
be rather long, but also due to several other important factors. It has
been suggested that a supermassive binary black hole in the nucleus
of the AGN may be responsible for both the existence of quasi-periodicity
(Lehto \& Valtonen 1996; Valtaoja et al. 2000) and the activity itself
(Sillanp\"a\"a 1999), as well as for structural changes on parsec scales
(Lobanov \& Roland 2002).  In this picture, activity can be triggered
by the periodic tidal perturbation of the primary black hole's accretion disk
by the orbital motion of a secondary black hole or the secondary's accretion
disk.  The eccentricity of the orbit (Lehto \& Valtonen 1996; Sundelius et al.
1997) may divide the overall period in two unequal parts, while precession of
the jet may change the viewing angle, resulting in modulation of the
amplitudes of outbursts. Relativistic precession of the secondary's
orbit will complicate the observed variability even more.

\begin{figure}
\hspace*{-1.0cm}
\includegraphics[width=0.45\textwidth,angle=-90]{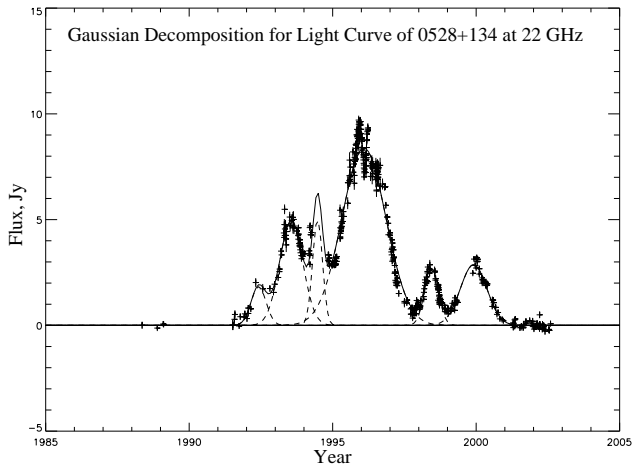}
\caption{Example of our Gaussian decompositions of the light curves:
light curve for 0528+134 at 22~GHz.}
\label{fig:gaussexam}
\end{figure}

In our approach, we carry out a joint analysis of the total-flux and spectral
variability based on multi-frequency light curves, combined with the observed
structural changes in VLBI images.  This paper presents our results for a
first small sample of four selected sources.
The frequency-dependent time delays found here can also be used to test models
of the nonthermal emission in blazars (Gomez et al. 1997; Lobanov 1998;
Marscher 2001).

\section{Description of the Data Used}

The combined data from the University of Michigan Radio Astronomy Observatory
(UMRAO; Aller et al. 1985) and Mets\"ahovi Radio Observatory (Ter\"asranta
et al.  1998, 2004, 2005) provide us with radio light curves for $\sim100$
AGN
from 4.8 to 37~GHz covering time intervals up to $\sim25$ years. A preliminary
inspection of all available light curves shows that only about one-third
display well separated, bright outbursts with definite frequency-dependent
time delays that are suitable for analysis.  As a first sample, we chose the
four $\gamma$-ray blazars 0458--020, 0528+134, 1730--130 and 2230+114, for
which multi-epoch VLBI data are also available (Jorstad et al. 2001). The 
multi-frequency light curves for these sources are shown in 
Figs.~2, 4, 7 and 9. In
all figures except for Fig.~\ref{fig:1730} (right) and \ref{fig:0528+2230vlbi}, 
filled symbols correspond to 
integrated measurements and hollow symbols to VLBI measurements; integrated 
measurements at 4.8~GHz are shown 
by diamonds, at 8~GHz by triangles, at 14.5~GHz by circles, 
at 22~GHz by squares, at 37~GHz by stars and at 90~GHz by pluses. 

\begin{figure*}
\centering
\begin{minipage}[c]{\textwidth}
\centering
   \includegraphics[width=0.47\textwidth]{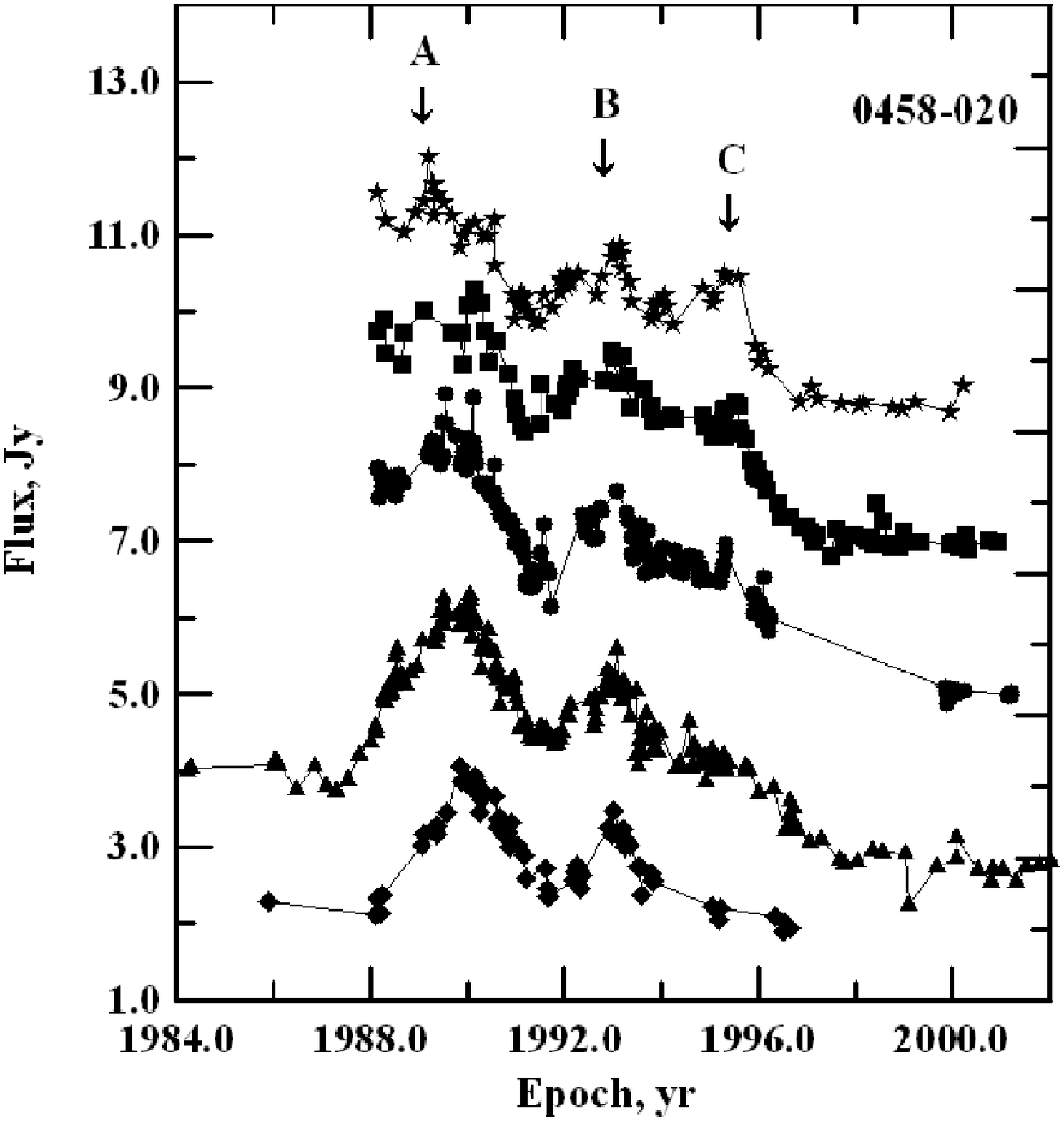}
   \hspace{0.02\textwidth}
   \includegraphics[width=0.47\textwidth]{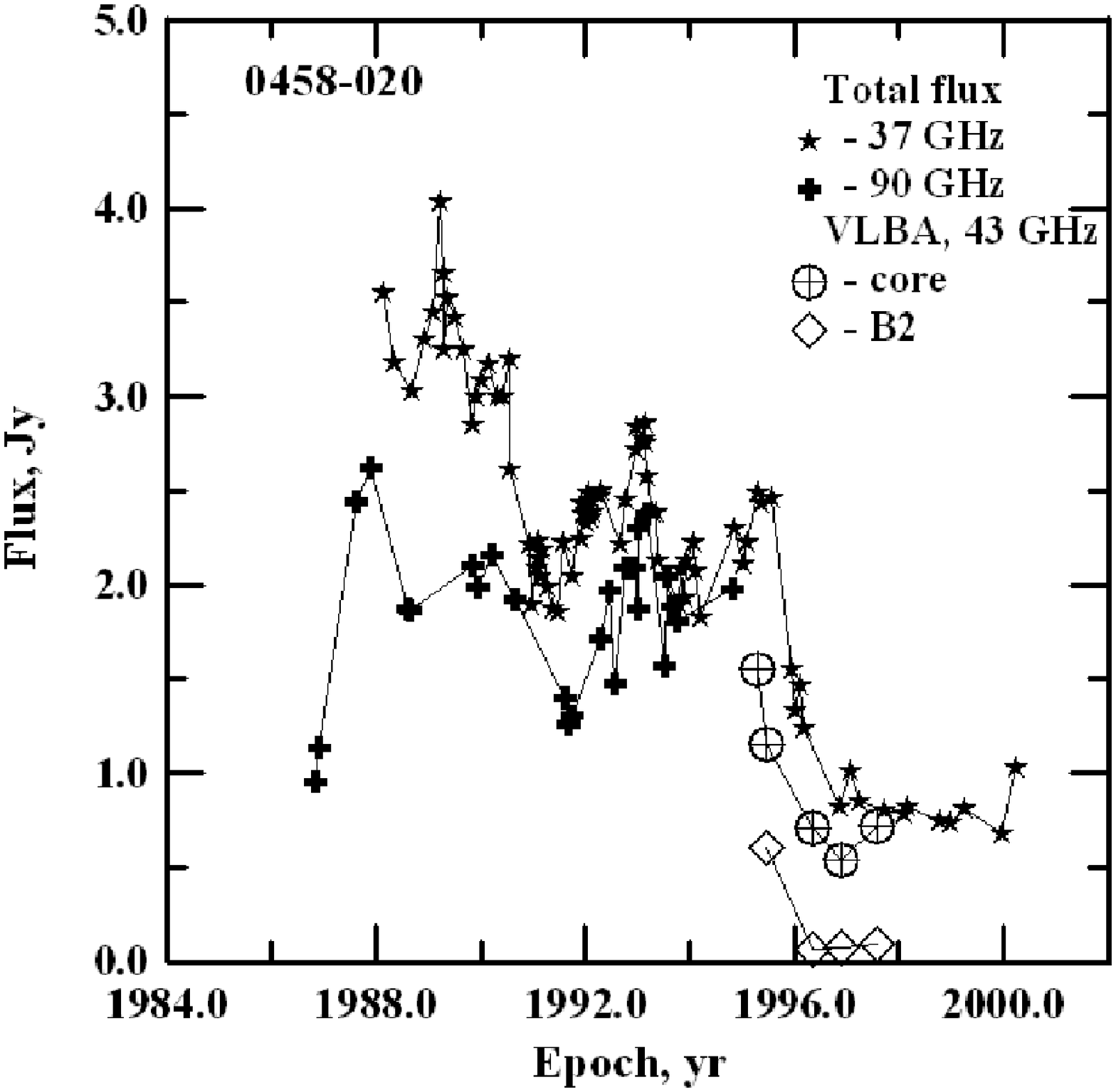}
\end{minipage}\\
\caption{{\bf 0458-020.} {\it Left}: From top to bottom, light curves 
for 37~GHz, 22~GHz, 14.5~GHz, 8~GHz, and 4.8~GHz. The light
curves have been shifted upward by 8~Jy, 6~Jy, 4~Jy, 1.5~Jy and 0~Jy,
respectively. {\it Right}: Light curves at 37 (filled stars) and 
90~GHz (filled pluses) combined with the flux-density variations of 
two components in the 43~GHz maps of Jorstad et al 2001.}
\label{fig:0458}
\end{figure*}
We separated the most prominent outbursts in the radio light curves
into individual components using Gaussian model fitting. We chose to fit
Gaussian components for the simple reason that most of the observed
centimetre-wavelength outbursts are symmetrical and have shapes that are
approximately Gaussian; in particular, the shapes of the centimetre-wavelength
outbursts do not correspond well to the exponential profiles used, e.g., by
Valtaoja et al. (1999) to fit millimetre flares. The highest of the five
frequencies for which we obtained light-curve decompositions is 37 GHz,
while the exponential behaviour pointed out by Valtaoja et al. (1999) is
manifest at frequencies of 90 and 230 GHz.

Long-term trends
in the total flux-density variations described using polynomial approximations
for the deepest minima in the light curves were subtracted before the fitting.
An example result of the long-term trend determination and Gaussian
decomposition is shown in Fig.~\ref{fig:gaussexam}. This example illustrates
the general criterion
that the smallest number of Gaussians providing a complete description of
the light curve was used. The number of Gaussians used depended on the
time interval covered by the light curve and the characteristic time scale
for the source variability.

The light curves of the sources considered here could all be decomposed into
Gaussian components with goodness of fit satifying the $\chi^2$ test at a
significance level of 0.01.
The frequency-dependent time delays for individual components,
$\Delta T_{max}(\nu)=T_{max}(\nu)-T_{max}(\nu_0)$, where $\nu_0$ is the highest
frequency observed, were determined and approximated by exponential functions
of frequency of the form $\Delta T_{max}(\nu)\propto\nu^{\alpha}$ (Gomez et al.
1997; Lobanov 1998). The results of these approximations are summarized in
Table~6.

The individual parameters of the outbursts as derived from the Gaussian
model fitting are given in Tables~1, 2, 3 and 5,
whose columns present (1) the component
notation, (2) the observing frequency, (3) the epoch of the maximum flux,
(4) the maximum amplitude, (5) the full-width at half maximum of the Gaussian
describing the outburst and (6) the time delay for the given frequency; when
the formal errors on the epoch of maximum, Gaussian width and time delay
were occasionally smaller than 0.01~yr, we set these errors equal to 0.01~yr,
Essentially all the Gaussian components fit are included in these tables;
the exception is that the Gaussians fit for the light curve of 1730-130
prior to $\simeq 1993$ were not included, since they all displayed very
small frequency delays and small amplitudes compared to the powerful
outburst of 1996.

\section{\bf{0458-020, $z=2.27$}}

\begin{figure}
\includegraphics[width=0.5\textwidth]{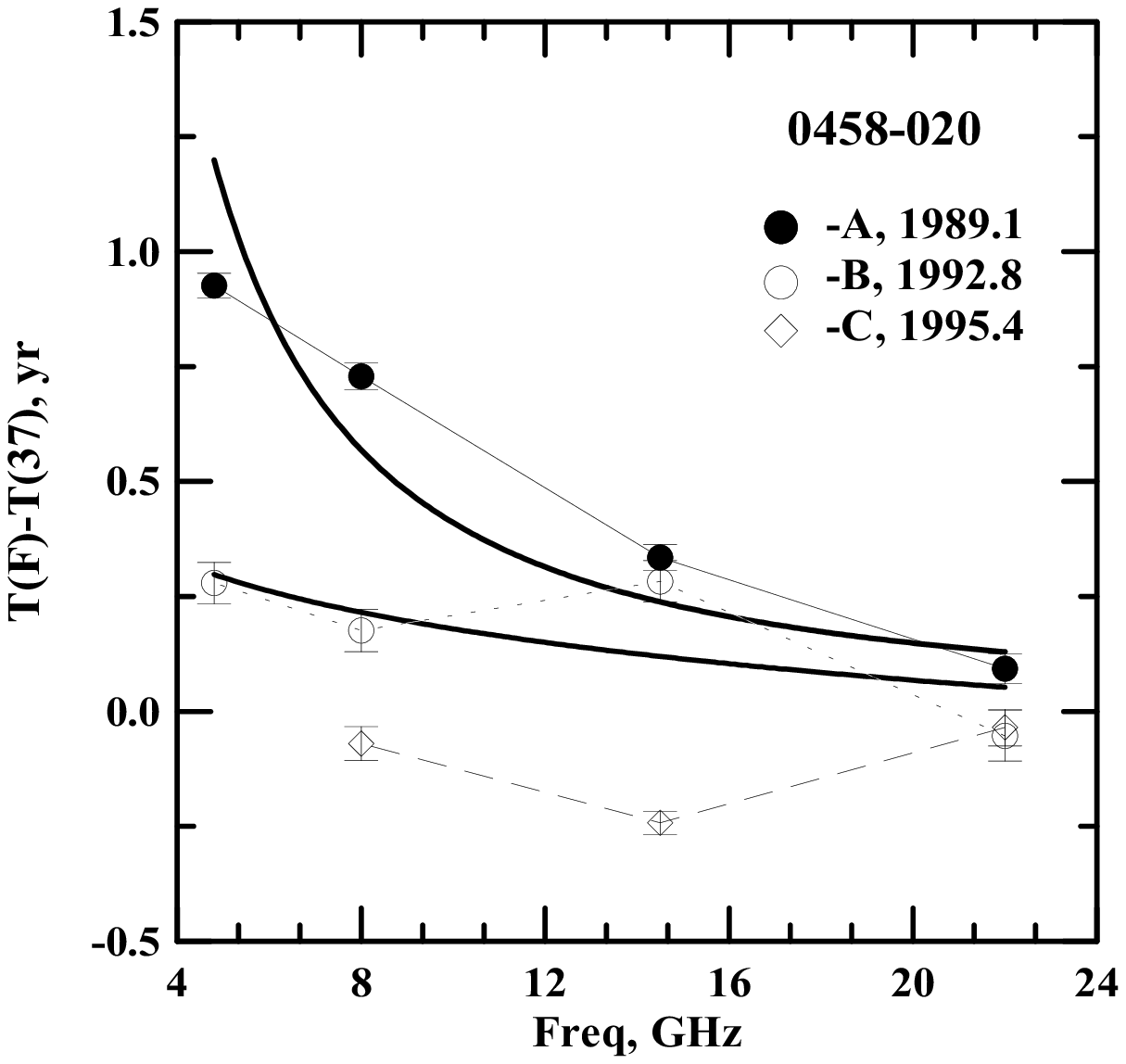}
\caption{{\bf 0458-020}. Time delays as functions of frequency. The bold
curves show exponential fits to the observed time delays.}
\label{fig:0458td}
\end{figure}

The radio light curve of this distant quasar displays three distinct outbursts
with declining amplitudes at epochs 1989.1, 1992.8 and 1995.4 in the period
1984--2002, which we label A, B and C, respectively (Fig.~\ref{fig:0458},
{\it left}). A
comparison of the 37~GHz light curve and the combined 90~GHz light curve based
on the data of Steppe et al. (1988, 1992, 1993) and Reuter et al. (1997)
indicates that at least outburst A (1989.06) may have some fine
structure that is unresolved in the available data (Fig.~\ref{fig:0458},
{\it right}).
The individual parameters of the outbursts as derived from the Gaussian
model fitting are given in Table~\ref{tab:0458}, and the
time delays for these outbursts as functions of frequency are shown
in Fig.~\ref{fig:0458td}.
Only outburst A displays frequency-dependent time delays, and may be
associated with a strong primary perturbation in the core.
Outbursts B (1992.81) and C (1995.40) show no time delays, and can accordingly
be considered jet outbursts. The evolution of the source's structure at 43~GHz
(Jorstad et al. 2001) shows the appearance of the new jet component B2
during 1995--1998. The variations of the 43~GHz flux densities of the core
and B2 shown in Fig.~\ref{fig:0458} ({\it right}) are consistent with the classification
of outburst C as a jet outburst.

\begin{table*}
\vspace{2mm}
\caption{\bf{0458-020} Parameters of outbursts}
\begin{center}
\begin{tabular}{llllll}
\hline
\noalign{\smallskip}
Comp. & Freq. & $T_{max}$     & Amplitude  & $\Theta$ & Time delay \\
      &  GHz  &  yr         &  Jy        &    yr    &   yr       \\
\noalign{\smallskip}
\hline
A & 37   & 1989.06$\pm$0.02 & 2.589$\pm$0.038 & 2.65$\pm$0.07 & 0             \\
  & 22   & 1989.16$\pm$0.02 & 3.292$\pm$0.074 & 2.82$\pm$0.08 & 0.09$\pm$0.03  \\
  & 14.5 & 1989.40$\pm$0.02 & 3.066$\pm$0.013 & 3.65$\pm$0.03 & 0.34$\pm$0.03 \\
  &  8   & 1989.79$\pm$0.02 & 2.465$\pm$0.010 & 2.83$\pm$0.02 & 0.73$\pm$0.03 \\
  &  4.8 & 1989.99$\pm$0.01 & 2.490$\pm$0.007 & 3.02$\pm$0.01 & 0.93$\pm$0.03 \\
B & 37   & 1992.81$\pm$0.04 & 1.700$\pm$0.034 & 3.23$\pm$0.15 & 0             \\
  & 22   & 1992.76$\pm$0.03 & 2.378$\pm$0.053 & 2.95$\pm$0.16 &-0.05$\pm$0.05  \\
  & 14.5 & 1993.09$\pm$0.01 & 2.225$\pm$0.018 & 1.54$\pm$0.02 & 0.28$\pm$0.05 \\
  &  8   & 1992.99$\pm$0.01 & 1.841$\pm$0.012 & 1.73$\pm$0.02 & 0.18$\pm$0.05  \\
  &  4.8 & 1993.09$\pm$0.01 & 1.780$\pm$0.013 & 1.57$\pm$0.01 & 0.28$\pm$0.04\\
C & 37   & 1995.41$\pm$0.02 & 1.295$\pm$0.054 & 1.11$\pm$0.06 & 0            \\
  & 22   & 1995.37$\pm$0.04 & 1.272$\pm$0.028 & 1.66$\pm$0.05 &-0.04$\pm$0.04  \\
  & 14.5 & 1995.17$\pm$0.02 & 1.639$\pm$0.012 & 1.94$\pm$0.02 &-0.24$\pm$0.02 \\
  &  8   & 1995.34$\pm$0.03 & 1.134$\pm$0.013 & 2.38$\pm$0.04 &-0.07$\pm$0.04  \\
\noalign{\smallskip}
\hline
\end{tabular}\end{center}
\label{tab:0458}
\end{table*}

Unfortunately, we have no high-resolution VLBI images for the period prior to
1995, and our classification of outbursts A (a core outburst) and B
(a jet outburst) cannot be confirmed by direct investigation of the structural
evolution. The only candidate for a jet component that could
be associated with outburst B is the
component B1 detected in 43~GHz maps at epochs 1995.31 and 1995.47.  The
radial distance of this component from the core is about 0.8~mas. If the rates
of angular separation from the core for components B1 and B2 are similar, the
zero-separation time for component B1 is near 1990. Therefore, there is some
possibility that the emergence and evolution of this component can be linked 
to outburst B.

We cannot identify the overall duration of the activity cycle in this
source $T_{act}$ because the available monitoring data spanning 16~years
include only one core outburst, and it is precisely the time interval
between core outbursts that determines $T_{act}$. The only thing we
can do at this stage is to give a lower limit for the total duration
of each activity cycle, $T_{act}\ge16$~yr, and to suggest that it appears
that the activity cycles in 0458-020 include a core outburst (as in
1989.06) followed by at least two jet outbursts (as in 1992.81 and 1995.41).

It is interesting that a comparison of the 15~GHz maps from the MOJAVE
database (Lister \& Homan 2005) for epochs 15 March, 2001 and 1 March, 2003
shows an increase in the peak flux between these two epochs, from
0.76 to 1.045~Jy. This brightening of the core could be an indication
of the beginning of a new activity cycle in this source.
Support for the onset of a new activity cycle can also be found
in the 43~GHz maps from the Radio Reference Frame Image Database
(RRFID; http://rorf/usno.navy.mil/rrfid.shtml). The peak flux density 
in the 43~GHz map for 13 September, 2003
exceeds the peak flux density in the map for 26 December, 2002 by more
than a factor of two ($S_{peak}=1.42$~Jy/beam as opposed to
$0.607$~Jy/beam).

If our lower limit for the characteristic durations of activity cycles in
0458--020 is correct, the maximum of the previous cycle should have
been sometime before 1973. Early monitoring data at 7.9~GHz (Dent and
Kapitzky 1976)
covering the time interval 1971--1974.5 display a moderate outburst with
its maximum flux, $S=2.86\pm0.09$~Jy, occurring near 1973. Some indications
of activity during the interval 1968--1979 can be found in the 5~GHz
monitoring data of Wright (1984) as well. These data show that 0458--020
was variable during this period, with a mean flux of $S_{mean}=1.76$~Jy
and an rms dispersion of 14\%. For comparison, the dispersion for the quasar
3C454.3, which displays a strong outburst during this period (UMRAO data),
was estimated to be 34\%. Thus, it seems probable that 0458--020 was
active during this period, although the level of activity
was probably not as strong as during the activity cycle observed in
roughly 1986--2002.

\section{\bf{0528+134}, $z=2.06$}

\begin{figure*}
\centering
\begin{minipage}[c]{\textwidth}
\centering
   \includegraphics[width=0.45\textwidth]{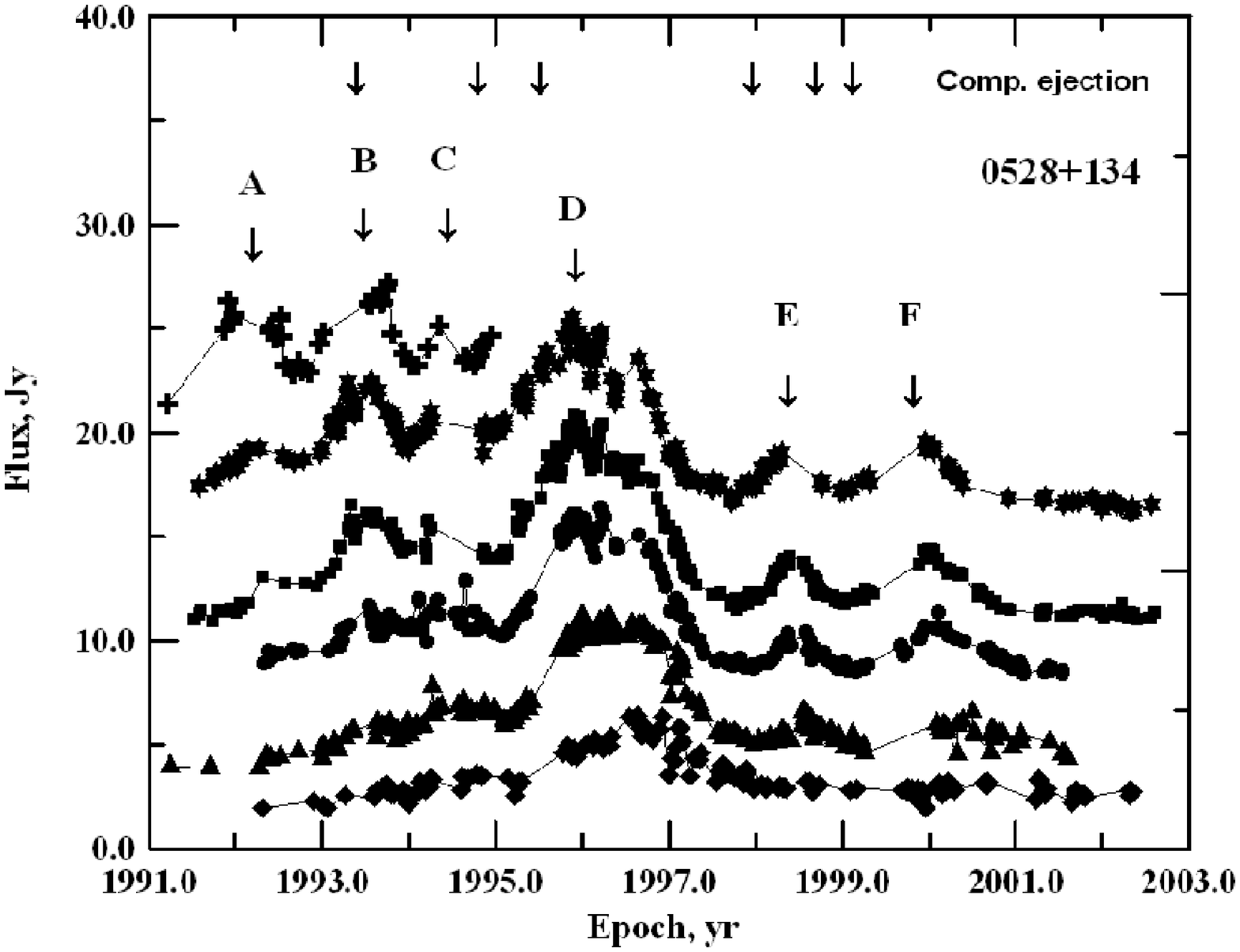}
   \hspace{0.02\textwidth}
   \vspace{0.2cm}
   \includegraphics[width=0.50\textwidth]{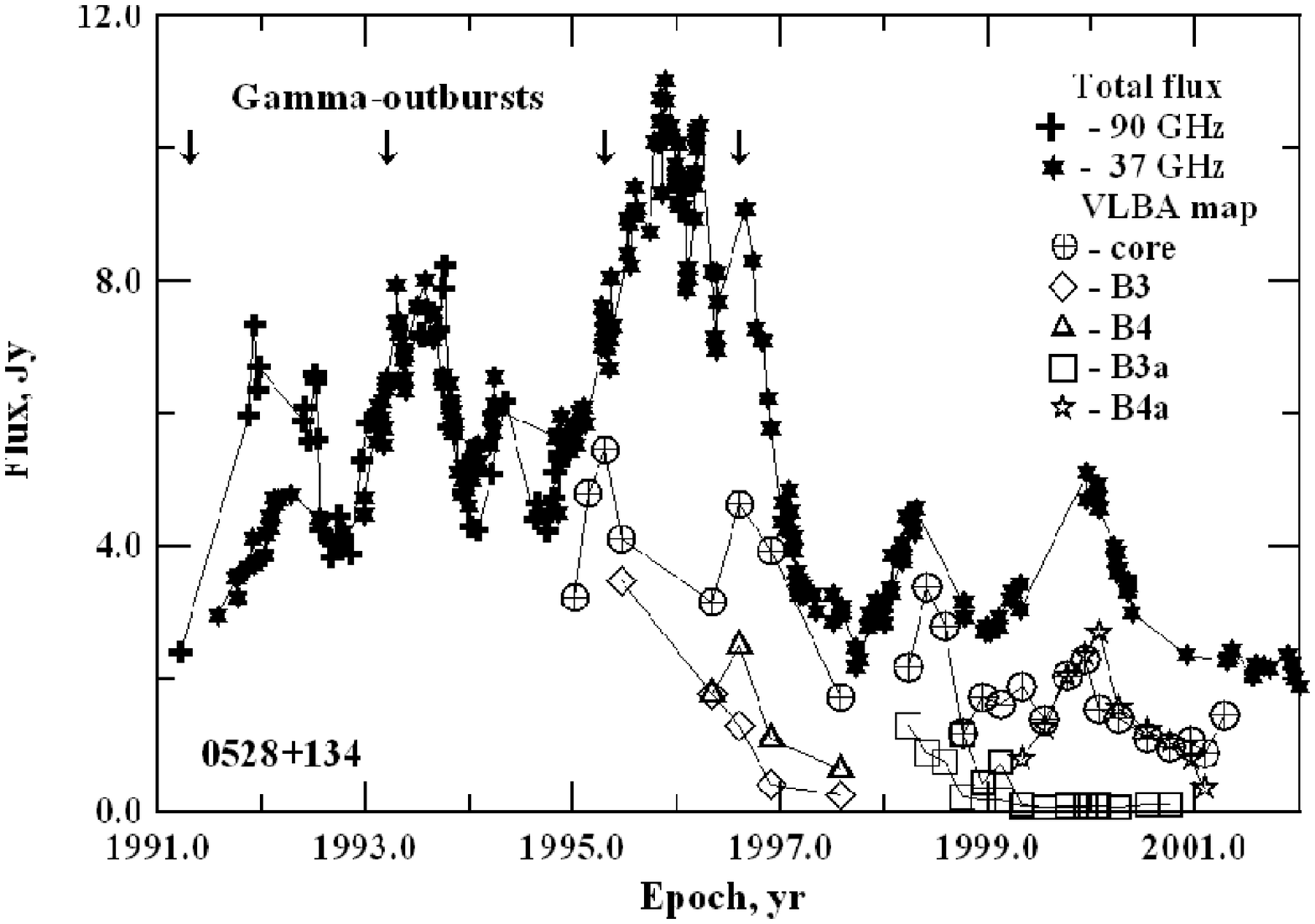}
\end{minipage}\\
\vspace*{-0.5cm}
\caption{{\bf 0528+134.} {\it Left}: From top to bottom, light curves for 
37~GHz, 22~GHz, 14.5~GHz, 8~GHz, and
4.8~GHz. The light curves have been shifted upward by
14.5~Jy, 9~Jy, 6~Jy, 2~Jy and 0~Jy, respectively.
{\it Right}: Light curves at 37 and 90~GHz combined with the flux-density
variations of individual components in the 43~GHz maps of Jorstad et
al. (2001, 2005).  The positions of bright $\gamma$-ray flares are shown
by the arrows.}
\label{fig:0528}
\end{figure*}

\begin{figure}
\includegraphics[width=0.50\textwidth]{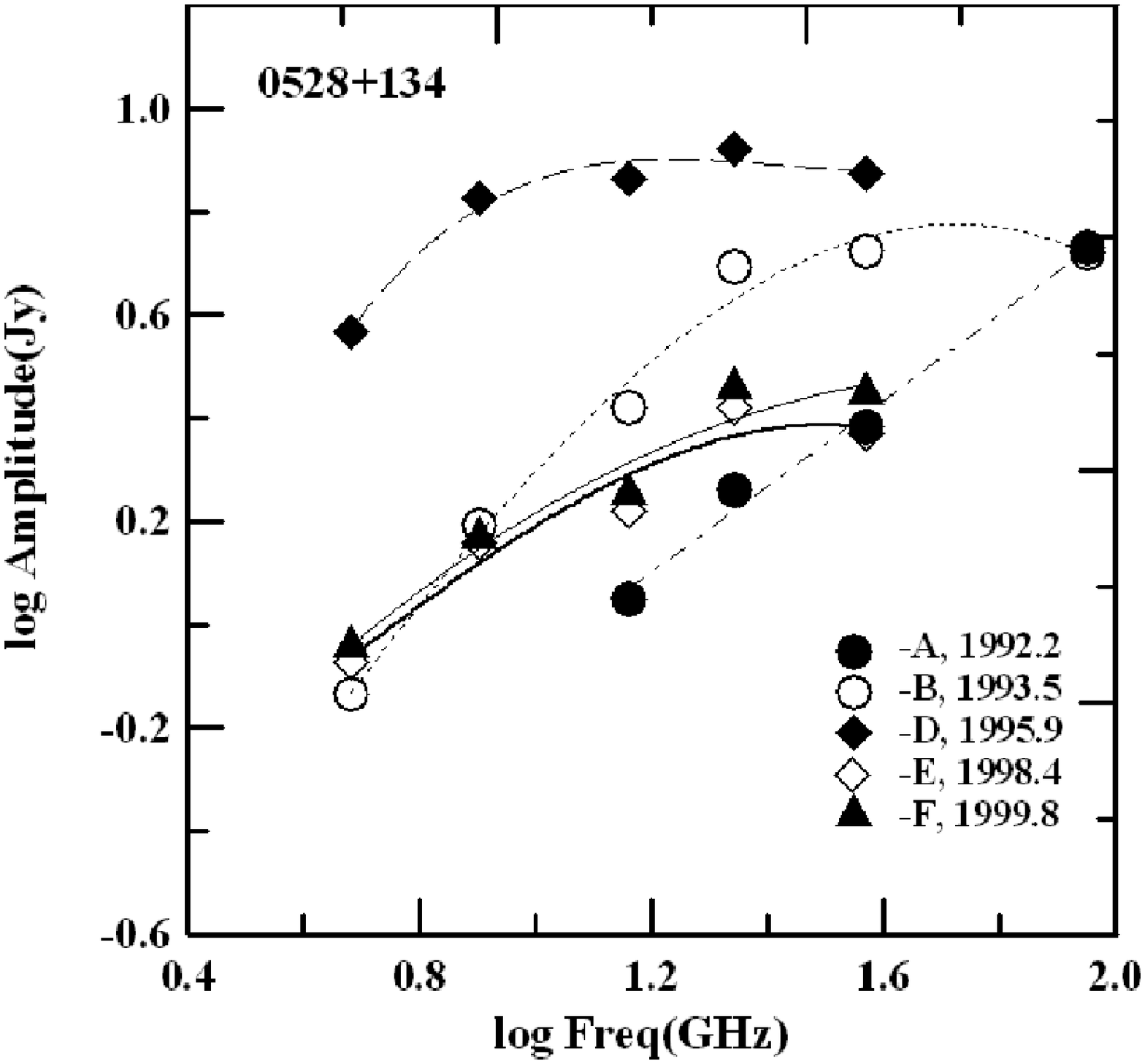}
\caption{{\bf 0528+134} Sub-outburst amplitudes as functions of frequency.
The curves show exponential fits to the observed amplitudes for sub-outbursts
A (dot--dash), B (dotted), D (dashed) and E/F (solid).}
\label{fig:0528sub}
\end{figure}

\begin{figure}
\includegraphics[width=0.5\textwidth]{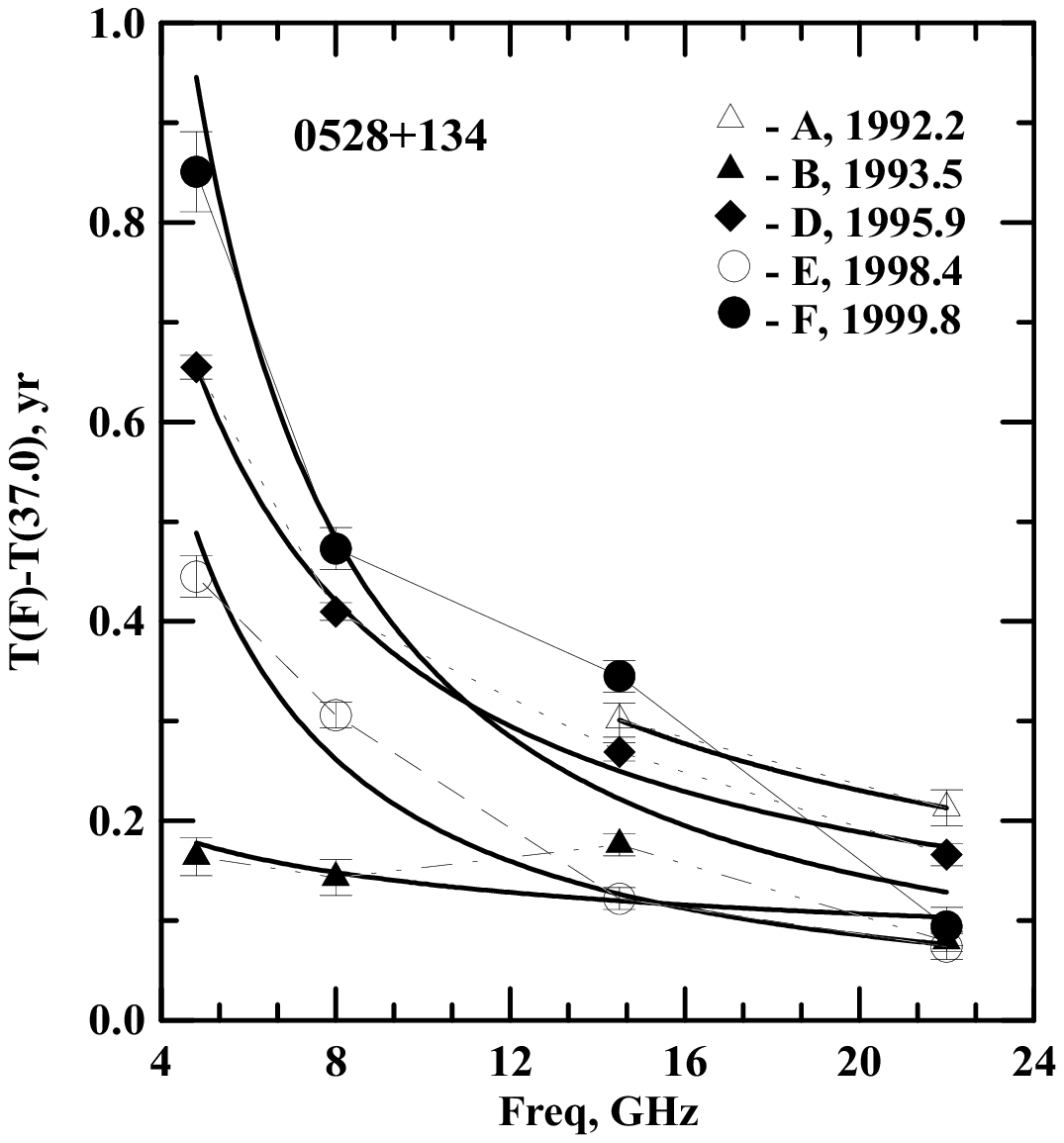}
\caption{{\bf 0528+134} Time delays as functions of frequency. The bold
curves show exponential fits to the observed time delays.}
\label{fig:0528td}
\end{figure}

The radio light curve of this quasar did not display bright events until
1990, when a strong and complex outburst started. The outburst displays at
least six narrow components (Fig.~\ref{fig:0528}, {\it left}), whose
parameters based on the Gaussian model fitting are given in
Table~\ref{tab:0528}. The 37~GHz light
curve shows the possible existence of additional fine structure in outburst D
that is unresolved by the Gaussian model fitting.
As was shown by Zhang et al. (1994), the spectrum of 0528+134 changed
dramatically at the end of the 1980s, when its turnover frequency moved
from near 8~GHz to near 100~GHz. A comparison of the 37~GHz (Mets\"ahovi)
and 90~GHz light curves
(see references above and Krichbaum et al. 1997) shows the
spectral evolution during this period.
The amplitudes
of individual outbursts as functions of frequency in Fig.~\ref{fig:0528sub}
show a shift of the turnover frequency in time, from above
90~GHz (A) to $\sim14.5$~GHz (D). In addition,
there is a sudden decrease in the amplitudes of outbursts E and F compared
to the earlier outbursts.  The
quasar 0528+134 is one of the brightest $\gamma$-ray sources.  The dates of
four EGRET flares with fluxes $>12\times
10^{-7}$phot$\cdot$cm$^{-2}\cdot$sec$^{-1}$ detected during 1991--1997
(Mukherjee et al. 1999)  are shown by the arrows in Fig.~\ref{fig:0528}
({\it right}).
The brightest flare (March 23--29, 1993; $T\sim1993.2$) coincides with
outburst B, while the brightest radio outburst (D) seems to be associated with
a moderate, prolonged $\gamma$-ray flare observed from April 4 to June 6, 1995
and a moderate flare detected in August 20--27, 1996. It is interesting that
the brightest $\gamma$-ray flare coincides with the brightest 90~GHz outburst,
B, but precedes the maximum of the brightest outburst at 37~GHz.  The
spectral evolution
combined with the $\gamma$-ray emission suggest that probably not all the
peaks observed in the 0528+134 light curves represent individual outbursts,
but are instead made up of a series of finer sub-outbursts that combine
to form what appears to be a single complex observed outburst. The parameters
of the Gaussian components fit at 90~GHz are very uncertain due to the
relatively poor sampling, and we adopt
37~GHz as a reference frequency for determining the time delays.  The delay
between 90 and 37~GHz averaged over sub-outbursts A, B and C is
$-0.006\pm0.045$. The frequency-dependent time delays for the sub-outbursts
are shown in Fig.~\ref{fig:0528td}. The time delays for the first sub-outburst, A
($\sim1992.1$), are determined only at $\nu\ge14.5$~GHz, since these flares
did not appear at the longer wavelengths.
The time delays at cm wavelengths are very small, even
for sub-outburst B (Fig.~\ref{fig:0528td}). Unfortunately, sub-outburst C ($\sim1994.4$)
coincides with a gap in the Mets\"ahovi data, and the time delays given
in Table~\ref{tab:0528} should be considered tentative.  The time delays of the later
sub-outbursts D, E and F vary over approximately a factor of two, from
0.444 to 0.850~yr.

High-frequency (22--86~GHz) VLBI monitoring during 1992--1997
(Zhang et al. 1994; Pohl et al. 1995; Krichbaum et al. 1995, 1997) has
revealed the appearance of five new jet components, which appear to be
associated with
sub-outbursts A (component N1, Krichbaum et al. 1995), B (component N2,
Krichbaum et al. 1995), C (component N3, Krichbaum et al. 1997) and D
(components N4 and N5, Krichbaum et al. 1997).  The multi-epoch imaging at
43~GHz of Jorstad et al. (2001, 2005) confirms the emergence of jet
components associated with sub-outbursts B (component B2), C (component B3)
and D (component B4), and has also revealed two new
jet components (B3a and B4a) associated with sub-outbursts E and F.
Thus, in contrast to the situation with the outbursts in 0458--020,
sub-outbursts A, B, C, D, E and F all show
signatures of both ``core'' outbursts (frequency-dependent time delays)
and ``jet'' outbursts (association
with the appearance of new superluminal components).
The time variations of the flux densities of the core and individual jet
components according to the data Jorstad et al. (2001, 2005) are shown
in Fig.~\ref{fig:0528} ({\it right}).

\begin{table*}
\begin{center}
\vspace{2mm}
\caption{\bf{0528+134} Parameters of outbursts}
\begin{tabular}{llllll}
\hline
Comp. & Freq. & $T_{max}$   & Amplitude  & $\Theta$ & Time delay \\
      &  (GHz)  &  (yr)        & (Jy)       &   (yr)   &  (yr)      \\
\hline
A & 90   & 1992.15$\pm$0.02 & 5.10$\pm$0.26 & 0.80$\pm$0.02 &-0.06$\pm$0.02\\
  & 37   & 1992.21$\pm$0.01 & 2.415$\pm$0.042 & 0.93$\pm$0.02 & 0 \\
  & 22   & 1992.42$\pm$0.01 & 1.83$\pm$0.17 & 0.62$\pm$0.04 & 0.21$\pm$0.02\\
  & 14.5 & 1992.51$\pm$0.01 & 1.124$\pm$0.043 & 0.63$\pm$0.06 & 0.30$\pm$0.02\\
B & 90   & 1993.51$\pm$0.01 & 5.12$\pm$0.11 & 0.90$\pm$0.01 & 0.03$\pm$0.02\\
  & 37   & 1993.48$\pm$0.01 & 5.310$\pm$0.050  & 0.97$\pm$0.01 & 0        \\
  & 22   & 1993.56$\pm$0.01 & 4.927$\pm$0.062 & 0.96$\pm$0.02 & 0.08$\pm$0.01\\
  & 14.5 & 1993.65$\pm$0.01 & 2.634$\pm$0.025 & 1.25$\pm$0.04 & 0.18$\pm$0.01\\
  &  8   & 1993.62$\pm$0.02 & 1.564$\pm$0.021 & 1.55$\pm$0.07 & 0.14$\pm$0.02\\
  &  4.8 & 1993.64$\pm$0.02 & 0.736$\pm$0.012 & 0.80$\pm$0.02 & 0.16$\pm$0.02\\
C & 90   & 1994.45$\pm$0.02 & 4.03$\pm$0.68 & 0.40$\pm$0.03 & 0.01$\pm$0.18\\
  & 37   & 1994.44$\pm$0.01 & 4.13$\pm$0.25 & 0.48$\pm$0.01 & 0             \\
  & 22   & 1994.47$\pm$0.01 & 5.01$\pm$0.19 & 0.41$\pm$0.01 & 0.03$\pm$0.01\\
  & 14.5 & 1994.52$\pm$0.01 & 2.311$\pm$0.031 & 0.64$\pm$0.01 & 0.08$\pm$0.01\\
  &  8   & 1994.52$\pm$0.01 & 1.921$\pm$0.040 & 0.59$\pm$0.01 & 0.07$\pm$0.01\\
  &  4.8 & 1994.55$\pm$0.01 & 1.329$\pm$0.023 & 0.69$\pm$0.01 & 0.11$\pm$0.01 \\
D & 37   & 1995.92$\pm$0.01 & 7.490$\pm$0.032 & 1.76$\pm$0.01 & 0             \\
  & 22   & 1996.09$\pm$0.01 & 8.385$\pm$0.023 & 1.81$\pm$0.01 & 0.16$\pm$0.01\\
  & 14.5 & 1996.19$\pm$0.01 & 7.327$\pm$0.021 & 1.56$\pm$0.01 & 0.27$\pm$0.01\\
  &  8   & 1996.33$\pm$0.01 & 6.725$\pm$0.017 & 1.90$\pm$0.01 & 0.41$\pm$0.01\\
  &  4.8 & 1996.58$\pm$0.01 & 3.694$\pm$0.013 & 1.97$\pm$0.01 & 0.65$\pm$0.01\\
E & 37   & 1998.37$\pm$0.01 & 2.338$\pm$0.073 & 0.62$\pm$0.01 & 0            \\
  & 22   & 1998.44$\pm$0.01 & 2.626$\pm$0.040 & 0.57$\pm$0.01 & 0.07$\pm$0.01\\
  & 14.5 & 1998.49$\pm$0.01 & 1.661$\pm$0.025 & 0.44$\pm$0.01 & 0.12$\pm$0.01\\
  &  8   & 1998.67$\pm$0.01 & 1.441$\pm$0.020 & 0.77$\pm$0.01 & 0.30$\pm$0.01\\
  &  4.8 & 1998.81$\pm$0.02 & 0.844$\pm$0.024 & 0.96$\pm$0.04 & 0.44$\pm$0.02\\
F & 37   & 1999.82$\pm$0.01 & 2.801$\pm$0.076 & 1.03$\pm$0.02 & 0            \\
  & 22   & 1999.91$\pm$0.01 & 2.874$\pm$0.031 & 1.09$\pm$0.01 & 0.09$\pm$0.02\\
  & 14.5 & 2000.16$\pm$0.01 & 1.788$\pm$0.012 & 0.93$\pm$0.01 & 0.34$\pm$0.02\\
  &  8   & 2000.29$\pm$0.02 & 1.474$\pm$0.012 & 1.36$\pm$0.02 & 0.47$\pm$0.02\\
  &  4.8 & 2000.67$\pm$0.04 & 0.905$\pm$0.011 & 2.04$\pm$0.04 & 0.85$\pm$0.04 \\
\hline
\end{tabular}
\end{center}
\label{tab:0528}
\end{table*}

\begin{figure*}
\centering
\begin{minipage}[c]{\textwidth}
\centering
   \includegraphics[width=0.425\textwidth]{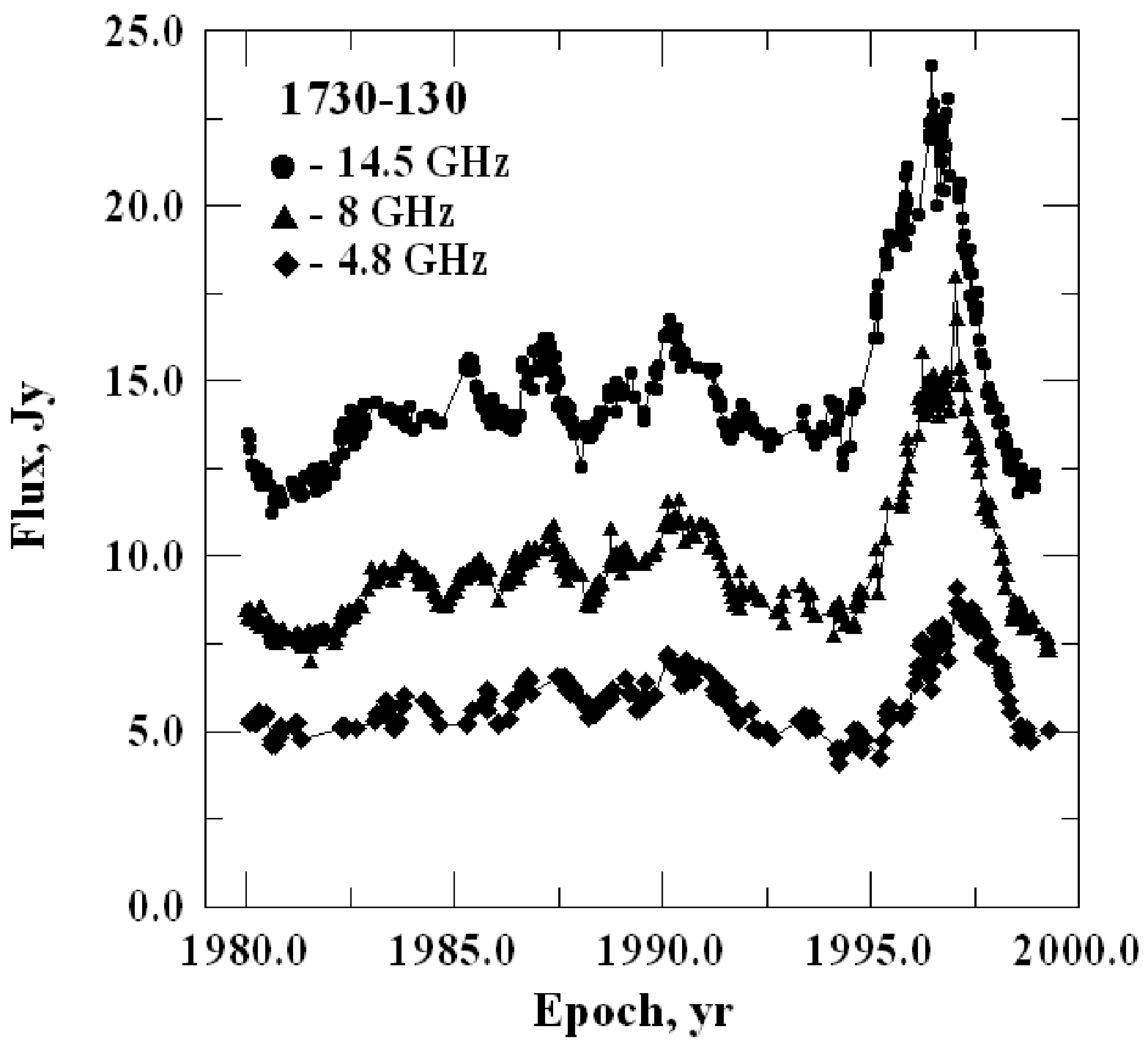}
   \hspace{0.02\textwidth}
   \vspace{0.1cm}
   \includegraphics[width=0.52\textwidth]{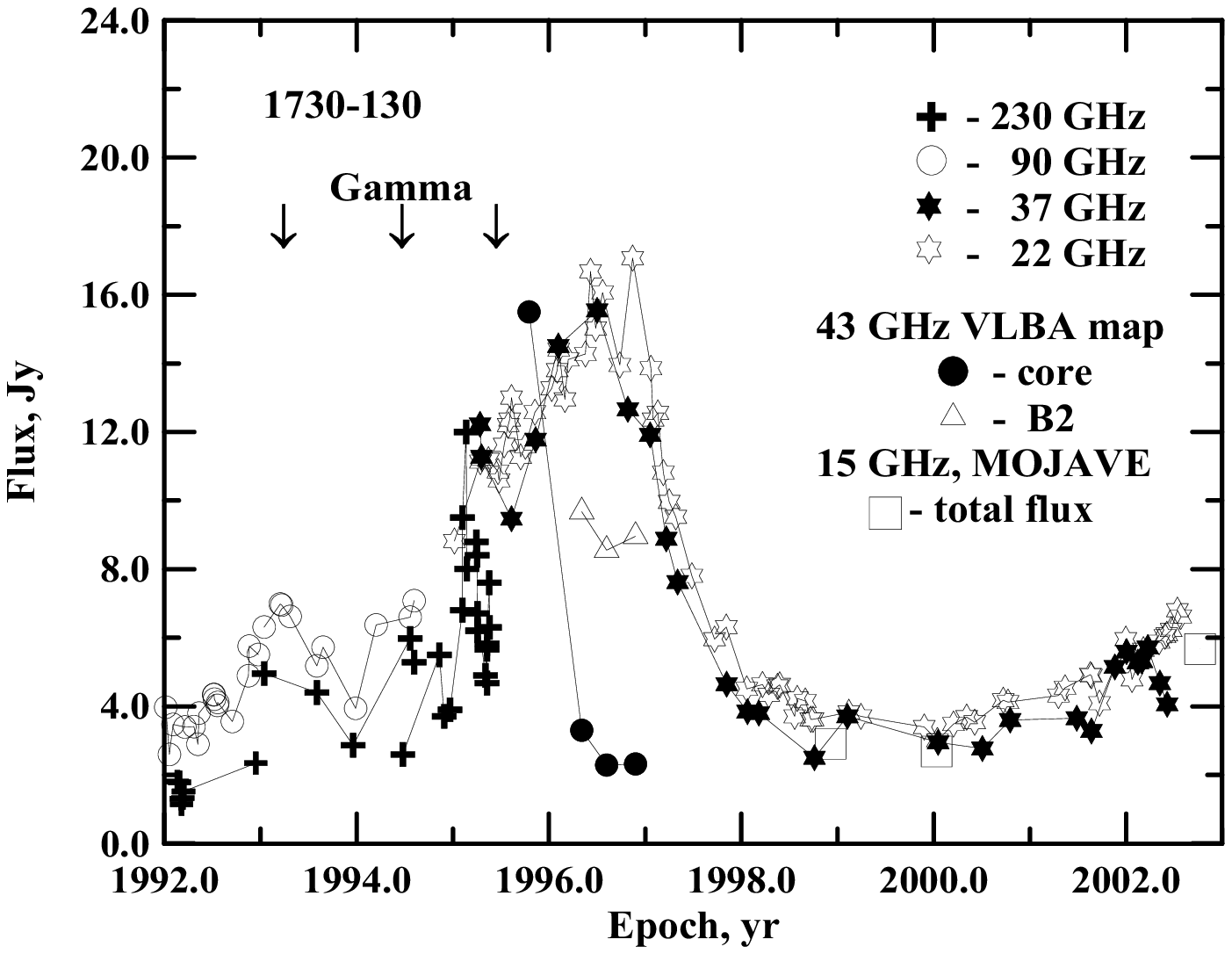}
   \end{minipage}\\
\vspace*{-0.5cm}
\caption{{\bf 1730-130} {\it Left}: From top to bottom, light curves for 
14.5~GHz, 8~GHz and 4.8~GHz shifted upward by
8~Jy, 3~Jy and 0~Jy, respectively. {\it Right}: Light curves at
22~GHz, 37~GHz (both Mets\"ahovi), 90~GHz and 230~GHz (Bower et al. 1997;
Steppe et al.1993; Reuter et al. 1997) combined with the flux-density
variations of two individual components in the 43~GHz maps of Jorstad et
al. (2001, 2005) and the total VLBI flux densities measured in the 
three MOJAVE images. The positions of $\gamma$-ray
flares are marked by the arrows.}
\label{fig:1730}
\end{figure*}

As we noted above, this outburst is the brightest event in this
object since 1976. However, low-amplitude flux-density variations were also
observed at cm wavelengths during the period of low activity (1977--1991).
Analysis of 8~GHz geodetic VLBI observations over nearly eight years
from 1986.25 to 1994.07 made by Britzen et al. (1999b) has revealed,
apart from the component N2 mentioned above (Krichbaum et al. 1997,
component {\bf a} in the notation of Britzen et al.), several superluminal
jet components ejected prior to 1990. It is interesting
that the position angles of the previously ejected components {\bf c}
($T_0=1984.4\pm2.3$, $PA\sim180^{\circ}$) and {\bf d} ($T_0=1982.1\pm2.3$,
$PA\sim150^{\circ}$) differ significantly from the position angle of
component {\bf a} ($T_0=1992.1\pm0.4$, $PA\sim70$) (Britzen et al 1999b).
Thus, it seems that we can identify different light curves for 0528+134
for two different periods characterized by low (1977--1991) and high
(1991--2004) activity.  The difference between
the observed levels of activity during these two periods could be due
to variations of the jet's orientation (Zhang et al. 1994) or some other
reasons internal to the source. It is interesting that the fine structure of
the 8~GHz outburst seems to be less prominent during the low-activity period.
The low level of activity and absence of high-frequency monitoring data
hinder the detection and comparison of fine structure of the outbursts
observed in the period of high and low activity.  We estimate
the characteristic duration of one activity cycle
in the source to be $\geq14$~yr.
Component {\bf b} (Britzen et al.  1999b), which was ejected during the
period of low activity ($T_0=1988.4\pm1.0$) in $PA\sim50^{\circ}$,
does not fit into the picture we have outlined.
Further high-frequency single-dish and/or VLBI observations are necessary
to test our proposed picture.

\section{\bf{1730-130 (NRAO 530), $z=0.90$}}

\begin{figure}
\includegraphics[width=0.5\textwidth]{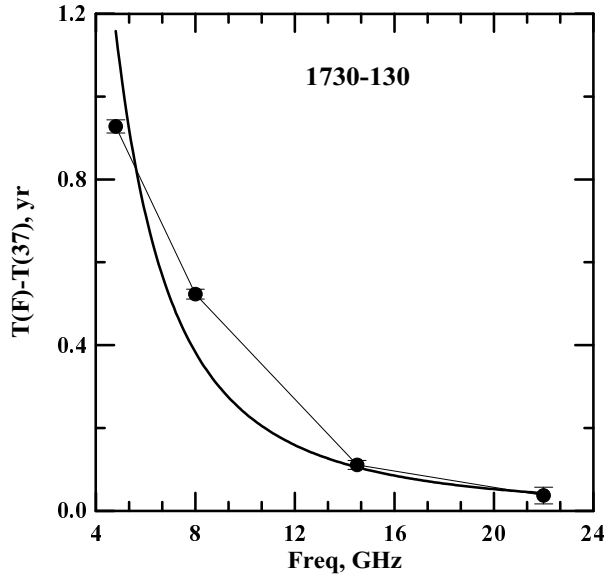}
\caption{{\bf 1730-130} Time delays as functions of frequency. The bold
curves show exponential fits to the observed time delays.}
\label{fig:1730td}
\end{figure}

This quasar has displayed a single bright outburst in $\sim$35 years
of cm-wavelength monitoring from 1967 to 2003 (Fig.~\ref{fig:1730},
{\it left}),
first detected as an extremely narrow and bright flare
at 230~GHz
(Bower et al. 1997). The combined 90~GHz and 230~GHz monitoring data
(Steppe et al.1993; Reuter et al. 1997) reveal two fainter outbursts
near 1993 and 1994.5, which probably precede the 230~GHz flare
(Fig.~\ref{fig:1730}, {\it right}) detected by Bower et al. (1997). All of the outbursts
coincide with epochs when $\gamma$-ray flares with fluxes
$>7\times10^{-7}$phot$\cdot$cm$^{-2}\cdot$sec$^{-1}$ were detected, marked in
the figure by arrows (Mukherjee et al. 1997). The brightest $\gamma$-ray flare
($\sim1995.45$) occurs slightly
after the main 230~GHz outburst, but precedes the main
maximum at mm wavelengths.  Comparison of the 90 and 37~GHz light curves
reveals some traces of fine structure, at least at high frequencies.
The parameters of the outburst are given in Table~\ref{tab:1730}, and the
time delays as
functions of frequency are shown in Fig.~\ref{fig:1730td}.

86~GHz VLBI maps obtained at epochs $\sim1993.3$ (Lobanov et al. 2000) and
1995.35 (Bower et al. 1997) show the core and two jet components that could
have been ejected during the 1993 and 1994.5 outbursts, but we do not
have enough information to be sure about this.
43~GHz VLBA imaging (Jorstad et al. 2001)
shows faint traces of a component that was probably ejected
prior to 1995 (B1, $T_0=1994.57\pm0.17$) and a new component that was
apparently ejected during the bright outburst (B2, $T_0=1995.54\pm0.05$).
The variations of the flux densities of the core and the new component B2 are
shown in Fig.~\ref{fig:1730} ({\it right}). Thus, the outburst exhibits signs of both
core and jet outbursts, like the outbursts observed in 0528+134.

\begin{table*}
\begin{center}
\vspace{-0.5cm}
\caption{\bf{1730-130} Parameters of outbursts}
\begin{tabular}{llllll}
\hline
Comp. & Freq. & $T_{max}$   & Amplitude  & $\Theta$ & Time delay \\
      & (GHz) & (yr)        & (Jy)       &   (yr)   &  (yr)      \\
\hline
A  & 37   & 1996.26$\pm$0.01  & 12.22$\pm$0.22 & 2.00$\pm$0.03 &  0       \\
   & 22   & 1996.30$\pm$0.02 & 11.28$\pm$0.17 & 2.11$\pm$0.03 & 0.04$\pm$0.02\\
   & 14.5 & 1996.37$\pm$0.01 & 10.139$\pm$0.031 & 2.09$\pm$0.01 & 0.11$\pm$0.01\\
   &  8   & 1996.78$\pm$0.01 &  7.091$\pm$0.020 & 2.03$\pm$0.01 & 0.53$\pm$0.01\\
   &  4.8 & 1997.19$\pm$0.01 &  3.379$\pm$0.017 & 2.02$\pm$0.02 & 0.93$\pm$0.02\\
\hline
\end{tabular}
\end{center}
\label{tab:1730}
\end{table*}

If we consider the 1993 and 1994.5 outbursts to be precursors of the main
outburst in 1996, the total duration of the activity cycle of the source
may correspond to the time interval from 1992 until 2001, when the
22 and 37~GHz fluxes began to grow again: $T_{act}\simeq 9$~yr. In this
case, the previous core outburst would have peaked in 1987, and the
UMRAO monitoring data (Fig.~\ref{fig:1730}, left) do show a modest
outburst at that epoch that seems to peak earliest at 14.5~GHz and latest
at 4.8~GHz (i.e., consistent with a core outburst).  Some further
support for this estimate for $T_{act}$ is provided in the MOJAVE
(15~GHz, Lister \& Homan 2005) and RRFID (24 and 43~GHz,
Fey et al. 2004) databases. The 15~GHz VLBI maps show that both the
total and peak flux densities grew by more than a factor of two between
January 11, 2000 and October 9, 2002, testifying to the onset of a new
outburst. The 15-GHz peak flux densities for October 9, 2002, February 11,
2004, March 23, 2005 and July 24, 2005 are 4.74, 3.33, 2.83 and
2.30~Jy/beam, respectively. This sequence can be supplemented
with the peak flux densities in the 24~GHz VLBI images reduced to the
beam size for the May 23, 2003 15-GHz observations: 3.7, 3.8 and
2.9 Jy/beam for May 16, 2002, August 26, 2002 and May 23, 2003. Overall,
these data clearly indicate a strong outburst in the VLBI core that seems
to have peaked in $\simeq 2002-2003$.  The decrease in the peak flux densities
at later epochs does not seem to be accompanied by the appearance of any
new structure, although there are clear changes in the polarization of
the core region between the first two 15-GHz epochs, overall suggesting that 
the 2002--2003 outburst may have been confined to the core.
It is possible that the 1996 outburst and 2002--2003 outburst are
associated with successive activity cycles in the source; if the 2002--2003
outburst is a ``precursor'' to a main core outburst occuring in 2005,
this could be consistent with our estimated cycle duration
$T_{act}\simeq 9$~yr (recall that the ``precursor'' outbursts for the
major 1996 outburst occured in 1993--1994), but further information about
the spectrum of the 2002--2003 outburst and further VLBI monitoring are
needed to test this hypothesis.

\section{\bf{2230+114 (CTA 102), $z=1.04$}}

\begin{table*}
\begin{center}
\vspace{2mm}
\caption{\bf{2230+114} Quasi-period in activity}
\begin{tabular}{lllllllll}
\hline
Freq &\multicolumn{2}{c}{Cycle 1}&\multicolumn{2}{c}{Cycle 2}&\multicolumn{2}{c}{Cycle 3}& P(2-1)& P(3-2)\\
     & Comp. & $T_{max}$ & Comp.& $T_{max}$ & Comp.& $T_{max}$ &   &  \\
37   &       &           &  A2   & 1989.57 &  A3  & 1997.48 &  & 7.91\\
22   &       &           &  A2   & 1989.67 &  A3  & 1997.90 &  & 8.22\\
14.5 &  A1   & 1981.36  &  A2   & 1989.74 &  A3  & 1998.00 & 8.39 & 8.25\\
8    &  A1   & 1981.40  &  A2   & 1989.95 &  A3  & 1998.34 & 8.55  & 8.39\\
37   &       &           &  B2   & 1990.50 &  B3  & 1997.92 &       & 7.42\\
22   &       &           &  B2   & 1990.54 &  B3  & 1998.57 &       & 8.02\\
14.5 &  B1   & 1982.71  &  B2   & 1990.63 &  B3  & 1998.62 & 7.92 & 7.98\\
8    &  B1   & 1982.94  &  B2   & 1990.84 &  B3  & 1999.09 & 7.89 & 8.25 \\
37   &       &           &  C2   & 1991.38 &  '3  & 1998.86 &       & 7.48\\
22   &       &           &  C2   & 1991.38 &  C3  & 1999.02 &       & 7.64\\
37   &       &           &  D2   & 1992.10 &  D3  & 2000.01 &       & 7.91\\
22   &       &           &  D2   & 1992.09 &  D3  & 1999.84 &       & 7.76\\
14.5 & C1+D1 & 1983.68  & C2+D2 & 1991.98 & C3+D3& 1999.75 & 8.3   & 7.76\\
8    & C1+D1 & 1983.97  & C2+D2 & 1991.71 & C3+D3& 2000.03 & 7.74 & 8.32\\
4.8  & C1+D1 & 1984.08  & C2+D2 & 1992.25 &      &          & 8.17 &  \\
\hline
\end{tabular}
\end{center}
\label{tab:2230quasi}
\end{table*}

\begin{figure*}
\centering
\begin{minipage}[c]{\textwidth}
   \centering
   \includegraphics[width=0.50\textwidth]{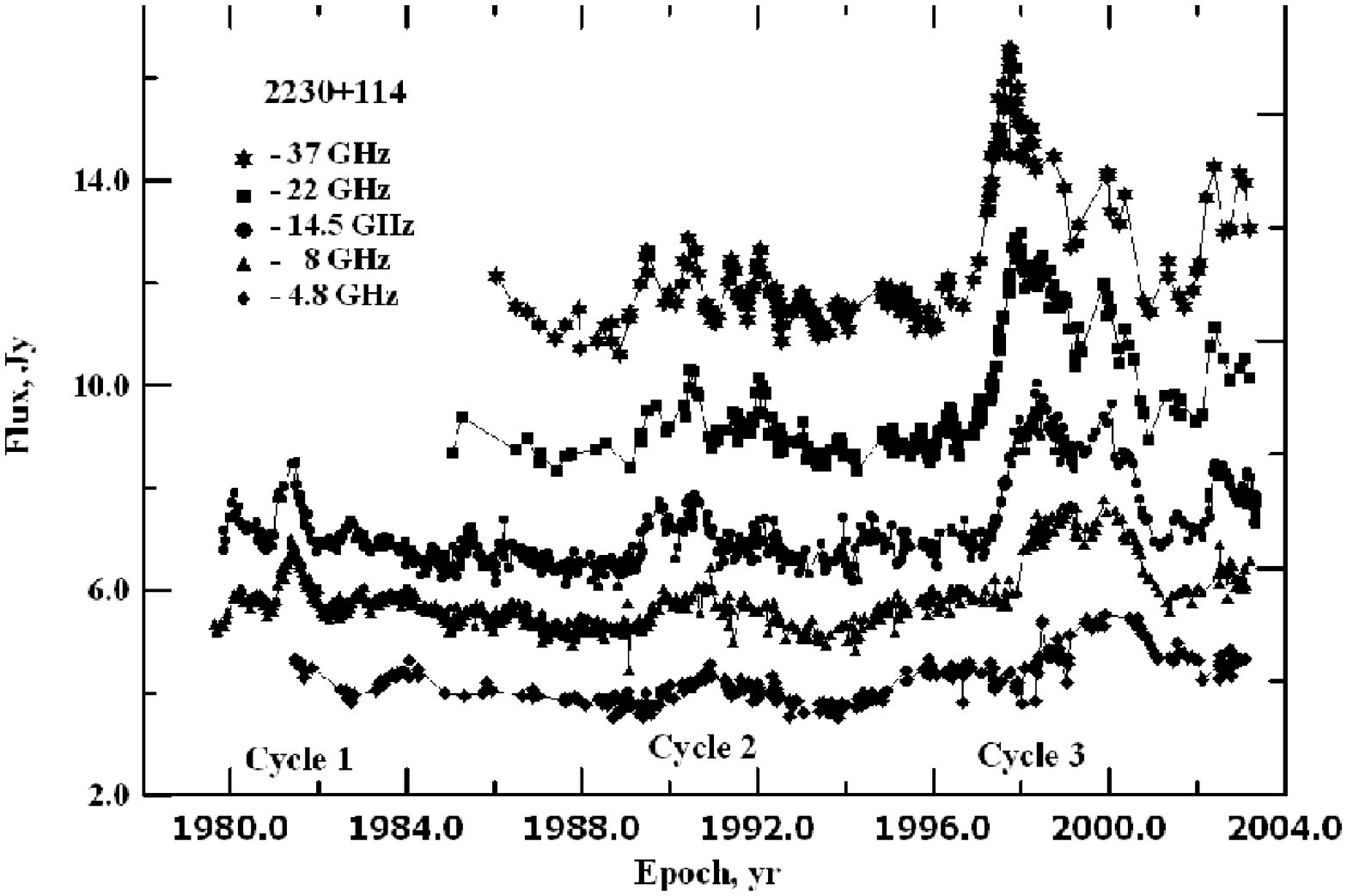}
\end{minipage}\\[-5truemm]
\centering
\begin{minipage}[c]{\textwidth}
   \centering
   \includegraphics[width=0.40\textwidth]{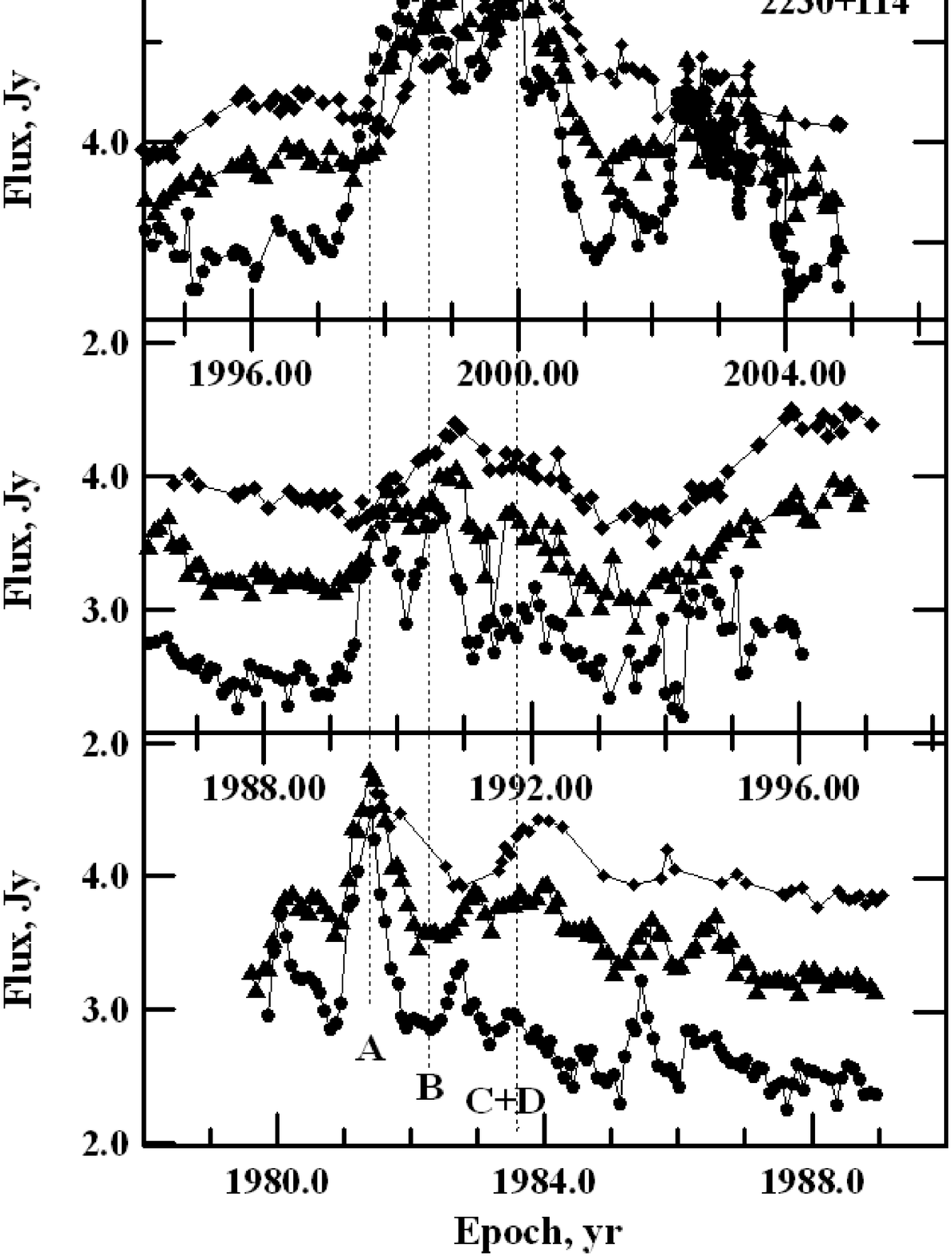}
   \hspace{0.02\textwidth}
   \includegraphics[width=0.40\textwidth]{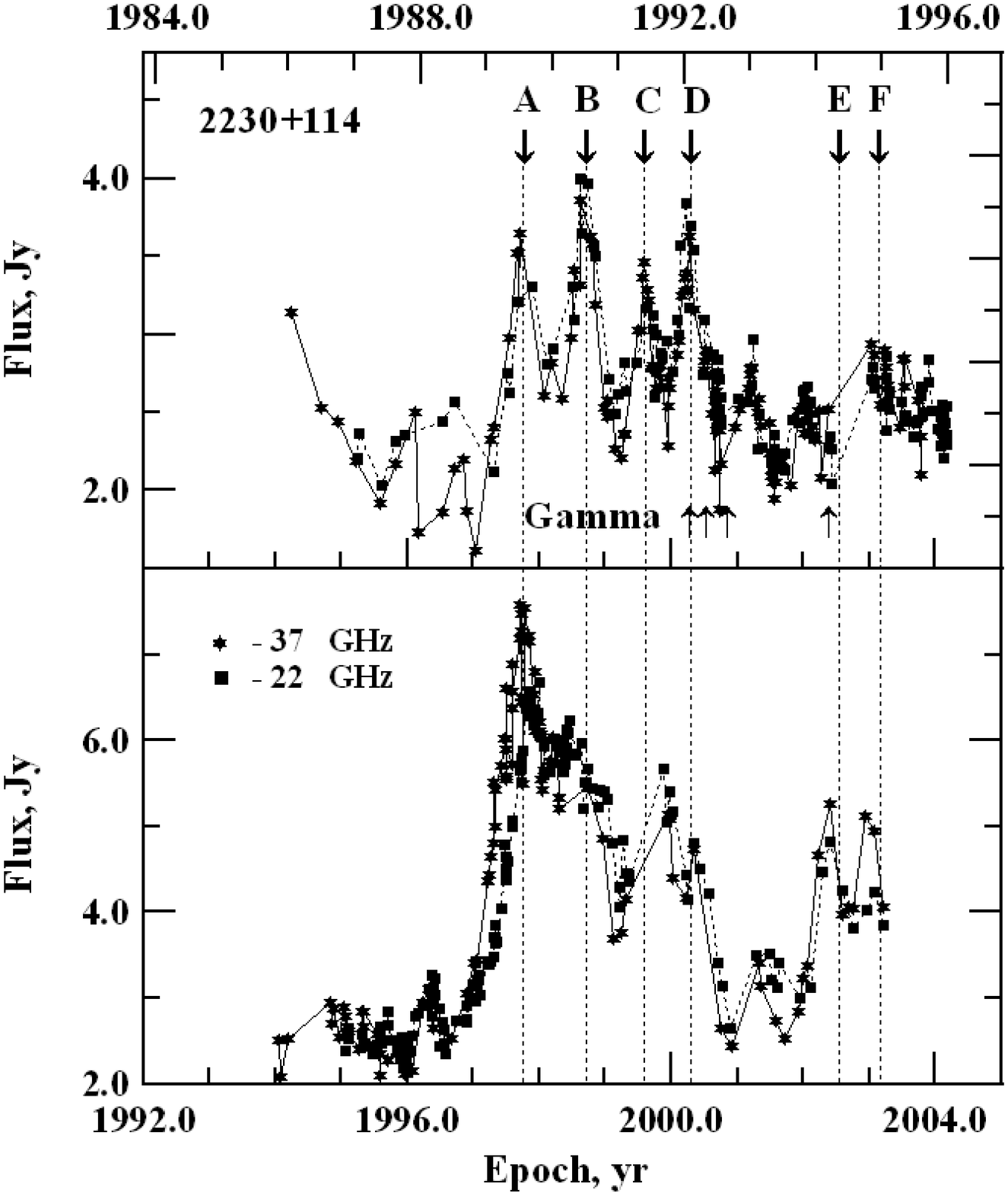}
\end{minipage}
\vspace*{-0.5cm}
\caption{{\bf 2230+114}. {\it Above}: From top to bottom, light curves 
for 37~GHz, 22~GHz, 14.5~GHz, 8~GHz and 4.8~GHz shifted upward by 
9~Jy, 6~Jy, 4~Jy, 2~Jy and 0~Jy,
respectively.  {\it Below left}: Quasi-periodicity of outbursts at
14.5~GHz (circles), 8~GHz (triangles) and 4.8~GHz (diamonds).
{\it Below right}: Quasi-periodicity at 37 and 22~GHz. In both the
lower figures, the years covered by each individual panel are indicated
beneath that panel, with some overlap between panels that cover adjacent
time intervals.  }
\label{fig:2230}
\end{figure*}

Radio light curves of this quasar display strong outbursts that repeat
approximately every $\sim$8 years (Fig.~\ref{fig:2230}).
At mm wavelengths, the outbursts can be decomposed into the six narrow
sub-outbursts A, B, C, D, E and F (Fig.~\ref{fig:2230}, {\it below
right});
only the two leading sub-outbursts A and B can be distinguished
at cm wavelengths, while the other sub-outbursts merge into two broad
features (C+D and E+F; Fig.~\ref{fig:2230}, {\it below left}).

To test whether there is quasi-periodicity in the appearance
of the outbursts, we subtracted the epochs for corresponding sub-outbursts
in consecutive cycles at all available wavelengths
(Table~\ref{tab:2230quasi}).  The mean
time intervals between the peaks of consecutive corresponding sub-outbursts
are $8.14\pm0.30$ for the interval from cycle 2 to cycle 1 and $7.95\pm0.31$
for the interval from cycle 3 to cycle 2. Given the time delays between the
various frequencies and the possible dependence of these time delays on the
amplitudes of individual sub-outbursts (see below), the agreement between
these two mean values is good.
Therefore, we tentatively suggest that outbursts in this source repeat every
$8.04\pm0.30$ years, preserving the relative positions in time of individual
sub-outbursts within the outbursts, even when the amplitudes of
the sub-outbursts change substantially.

In this case, we expect that the leading sub-outburst of the next outburst
reached its maximum at 37~GHz near $2005.5\pm0.3$, and its maximum at
$\simeq 15$~GHz near $2006.0\pm 0.3$. The MOJAVE VLBI images clearly
indicate an outburst in the VLBI core starting sometime in the second
half of 2004: the observed peak flux densities for successive epochs
are 1.05 (April 10, 2004), 1.87 (February 5, 2005), 2.00 (September 19, 2005),
3.94 (December 22, 2005) and 4.40~Jy/beam (February 12, 2006). This may well
represented the predicted leading sub-outburst of the next outburst cycle,
but further VLBI and integrated monitoring data are needed to determine the
epoch of the maximum flux and the spectrum of this outburst.
The leading 8~GHz maximum of the outburst that took place prior
to 1981 should have occured near 1973.5. However, 8~GHz monitoring data
(Dent \& Kapitzki 1976) show only slow flux variations with
an amplitude of $S\le1$~Jy during 1970.5--1974.5. If our hypothesis
of quasi-periodicity in the activity of the source
is correct, then the weak activity observed during the 1970s
testifies that the activity as a whole is strongly modulated in amplitude
by some geometric effect (e.g. variation of the viewing angle) or some
intrinsic reason.

Owing to the merging of sub-outbursts C, D and E, F at cm wavelengths, time
lags were determined only for the two main sub-outbursts A and B
(Table~\ref{tab:2230},
Fig.~\ref{fig:2230}). 2230+114 is the only source for which we were able to analyze outbursts
that may belong to different activity cycles. The time lags for the two
periods of activity are correlated with the corresponding outburst amplitudes:
sub-outbursts A and B of the brightest outburst, with its maximum near 1997.5,
show greater time lags than sub-outbursts A and B of the more modest
outburst with its maximum near 1989.6 (see below).

Four moderate $\gamma$-ray flares
($S>7\times10^{-7}$phot$\cdot$cm$^{-2}\cdot$sec$^{-1}$)
detected during 1992--1994 coincide with sub-outbursts D2 and
E2 (Fig.~\ref{fig:2230}, {\it below right}).
Unfortunately we have no high resolution VLBI data for epochs prior to 1995.
Multi-epoch 43-GHz VLBA observations (Jorstad 2001, 2005)
covering the time interval from 1995 to 2001 reveal three faint ($S\le0.3$~Jy)
superluminal components, which could have been ejected during the final stage
of cycle 2 (B1, $T_0=1994.12$; B2, $T_0=1995.19$; B3, $T_0=1996.08$), as well
as the two superluminal components B2a and B3a, which seem to be associated with
sub-outbursts B3 and C3. The variations of the flux densities of the core and
new-born jet components are shown in Fig.~~\ref{fig:2230vlbi}.
The frequency-dependent-delay sub-outburst A can be classified as a 
``core'' outburst, and is associated with brightening of the core. 
Sub-outburst B
is a mixed outburst, which displays low-frequency delays but is also
associated with the appearance of new superluminal components.

\begin{table*}
\begin{center}
\vspace{2mm}
\caption{\bf{2230+114} Parameters of outbursts}
\begin{tabular}{llllll}
\hline
Comp. & Freq. & $T_{max}$   & Amplitude  & $\Theta$ & Time delay \\
      &  GHz  &  yr         &  Jy        &    yr    &   yr       \\
\hline
A2 &  37   & 1989.57$\pm$0.01& 1.649$\pm$0.050 & 0.74$\pm$0.03&      0    \\
   &  22   & 1989.67$\pm$0.02& 1.21$\pm$0.11& 0.68$\pm$0.08& 0.10$\pm$0.02\\
   &  14.5 & 1989.74$\pm$0.01& 1.328$\pm$0.026& 0.70$\pm$0.03& 0.17$\pm$0.01\\
   &   8   & 1989.95$\pm$0.02& 0.730$\pm$0.024& 0.80$\pm$0.04& 0.38$\pm$0.02\\
   &   4.8 & 1989.94$\pm$0.02& 0.318$\pm$0.040 & 0.40$\pm$0.19& 0.37$\pm$0.02\\
A3 &  37   & 1997.48$\pm$0.01& 3.656$\pm$0.058& 0.57$\pm$0.08&      0   \\
   &  22   & 1997.90$\pm$0.01& 3.831$\pm$0.030 & 0.83$\pm$0.03& 0.42$\pm$0.01\\
   &  14.5 & 1998.00$\pm$0.01& 2.284$\pm$0.024& 0.90$\pm$0.02& 0.52$\pm$0.01\\
   &   8   & 1998.34$\pm$0.01& 1.411$\pm$0.023& 0.67$\pm$0.05& 0.86$\pm$0.01\\
B2 &  37   & 1990.50$\pm$0.01& 1.736$\pm$0.050 & 0.55$\pm$0.02&      0   \\
   &  22   & 1990.54$\pm$0.01& 1.59$\pm$0.10& 0.62$\pm$0.05& 0.04$\pm$0.01\\
   &  14.5 & 1990.63$\pm$0.01& 1.418$\pm$0.027& 0.66$\pm$0.02& 0.13$\pm$0.01\\
   &   8   & 1990.84$\pm$0.01& 0.965$\pm$0.021& 0.71$\pm$0.03& 0.33$\pm$0.01\\
   &   4.8 & 1990.92$\pm$0.03& 0.702$\pm$0.014& 1.34$\pm$0.08& 0.42$\pm$0.03\\
B3 &  37   & 1997.92$\pm$0.01& 3.630$\pm$0.059& 0.55$\pm$0.03&      0   \\
   &  22   & 1998.57$\pm$0.01& 2.83$\pm$0.11& 0.48$\pm$0.06& 0.64$\pm$0.01\\
   &  14.5 & 1998.62$\pm$0.01& 2.109$\pm$0.018& 0.66$\pm$0.02& 0.69$\pm$0.01\\
   &   8   & 1999.09$\pm$0.01& 1.766$\pm$0.026& 0.69$\pm$0.02& 1.16$\pm$0.01\\
\hline
\end{tabular}
\end{center}
\label{tab:2230}
\end{table*}

\begin{figure}
\includegraphics[width=0.50\textwidth]{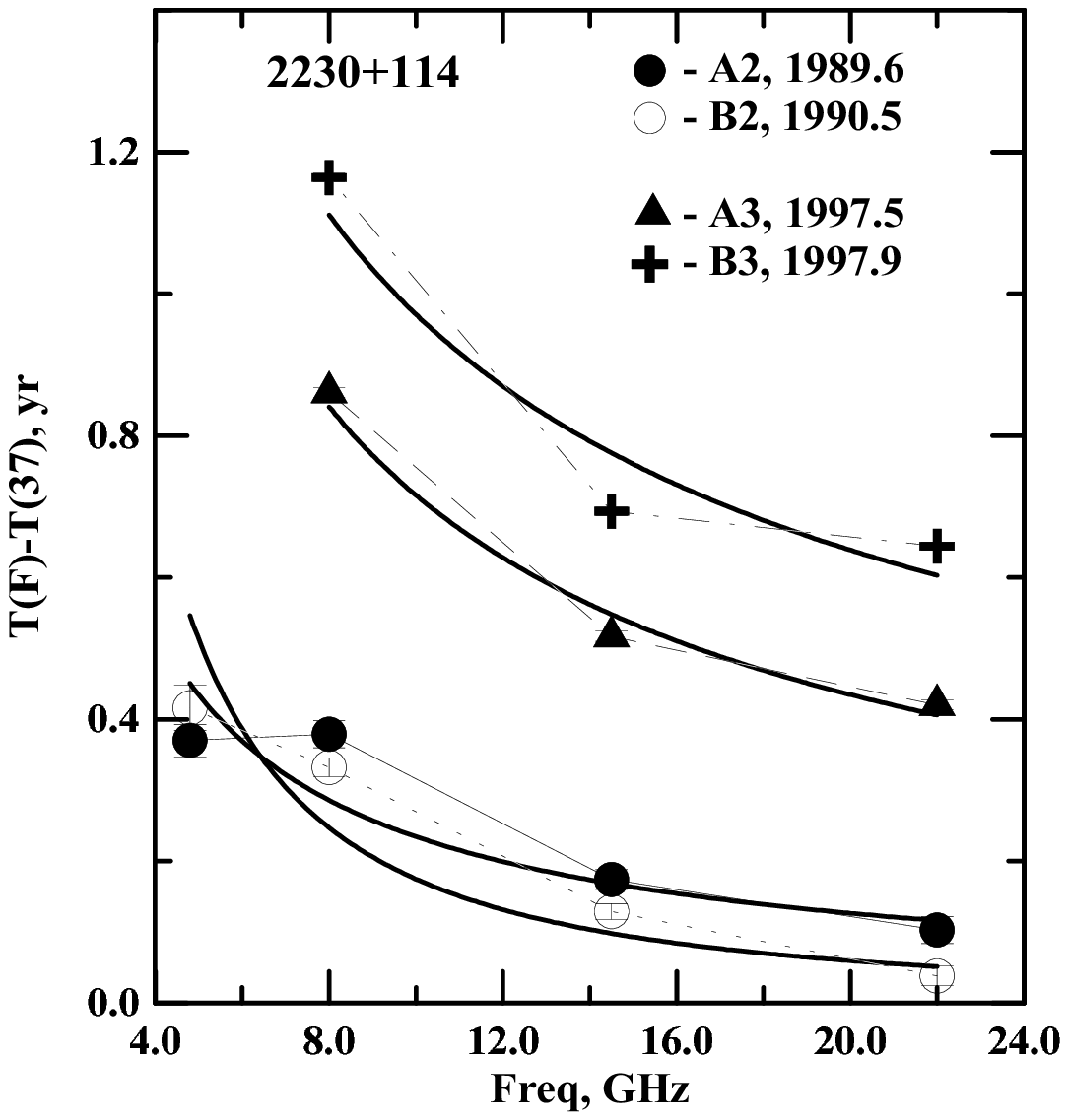}
\caption{{\bf 2230+114} Time delays as functions of frequency. The bold
curves show exponential fits to the observed time delays.}
\label{fig:2230td}
\end{figure}

\begin{figure}
\includegraphics[width=0.5\textwidth]{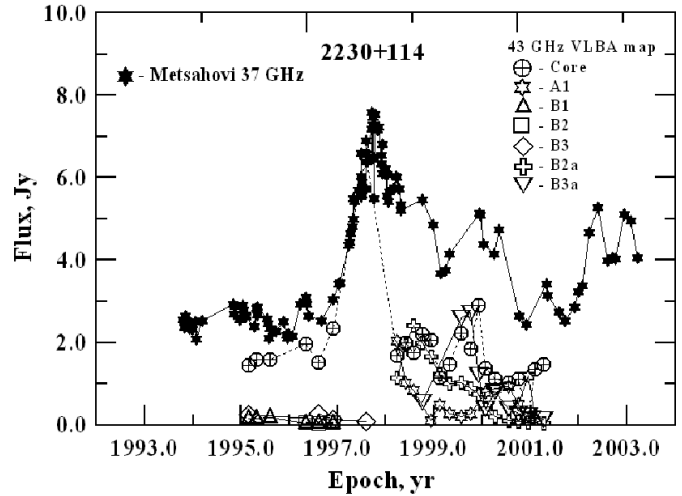}
\caption{{\bf 2230+114} Evolution of the 43~GHz flux densities of the core
and jet components (Jorstad et al. 2001, 2005)}
\label{fig:2230vlbi}
\end{figure}

\section{Discussion}

\subsection{Types of outbursts}

Analysis of the outbursts combined with the detailed structural evolution at
43~GHz shows that only one of the outbursts can be classified as a pure core
(2230+114, A) and one as a pure jet (0458-020, C) outburst in terms of both
their time delays and their association with core or jet brightening. Most
outbursts are mixed, in the sense that they display both frequency-dependent
time lags and a connection with the appearance of new jet components.
Note that previous classifications of outbursts (Pyatunina
et al. 2000; Zhou et al. 2000) were based on low frequency (5 and 8~GHz)
VLBI observations.
However, a jet component resolved at 43~GHz would be merged with the core at
lower frequencies, so that an outburst classified as a core type at low
frequency might be identified as a mixed type at high frequencies. This
illustrates the importance of high frequency VLBI data for probing the early
stages of activity within the jet.

It seems likely that the pure core and mixed core--jet outbursts represent
fundamentally the same phenomenon: activity in the core that gives rise to
the ejection of material along the jet. In this case, the brightness of the
ejected material associated with the pure core outburst A in 2230+114
faded before becoming
distinct from the observed core at 43 GHz. It is easy to understand the larger
number of pure core outbursts at lower frequencies as well: with the lower
resolution available, the ejected components are more likely to have faded
beyond detection by the time they become distinct from the core. The relatively
small number of pure jet outbursts then represent brightening of an optically
thin region of the jet well after its ejection from the core, e.g., due to the
formation of shocks.

\subsection{Fine structure of outbursts}

Some traces of fine structure can be found in all of the outbursts, at least
at the highest frequencies. In some of the sources, the fine structure
dominates, splitting outbursts into distinct sub-outbursts (0528+134,
2230+114), while in others (0458-020 and 1730-130), it degenerates into small
fluctuations of the fluxes. Note that the
appearance of fine structure in light curves is independent of resolution,
but can be hidden due to high optical depth.
Savolainen et al. (2002) suggested that outburst
fine structure can be induced by shocks that grow and decay in the innermost
few tenths of a milliarcsecond.
According to theoretical simulations, shocks can be
generated by several mechanisms: (a) multiple
recollimation shocks, which are induced by the pressure mismatch between the
jet and surrounding medium (Gomez et al. 1997, 2003);
(b) interactions between helical instability modes that grow
due to perturbations to the jet by jet--cloud collisions or precession,
with elliptical (Lobanov, Hardee \& Eilek 2003) or body waves
(Hardee et al. 2001) growing into the jet; and
(c) interactions of the jet with obstacles in its path. The preservation
of the positions of individual sub-outbursts in outbursts belonging
to different activity cycles in 2230+114 may provide important information
for distinguishing between these mechanisms. However, it is necessary to carry
out similar analyses for more sources, in order to investigate whether
this property is common or peculiar to this source.

The presence of fine structure in the outbursts can significantly complicate
the identification of a ``starting point'' of the activity. Strictly speaking,
only information about flux variations at frequencies above
$\sim40$~GHz in combination with 43~GHz VLBI data can address this problem.

\subsection{Time delay as a function of frequency}

The parameters of the time-delay function $\Delta T=\Delta T_0\nu^{\alpha}$
for all events defined for at least four frequencies are given in
Table~\ref{tab:summary}.
According to theoretical estimates (Gomez et al. 1993, 1997), the position of
the core
is expected to depend on
the frequency as $r_{core}\propto\nu^{-0.8}$. The frequency-dependent
shift of the core position in 3C345 estimated by Lobanov (1998) turned out
to be close to the theorical expectation $r\propto\nu^{(-1.04\pm0.16)}$.
If the flow velocity is independent of frequency, then the time delay
is expected to behave similarly. The indices of the time-delay functions
given in Table~\ref{tab:summary}
vary from --0.6 (0528+134, B3) to --2.2 (1730--130) with
the mean value $\alpha_{mean}=(-1.20\pm0.49)$. The time-delay functions for
individual events (Figs.~\ref{fig:0458td}, \ref{fig:0528td}, \ref{fig:1730td}
and \ref{fig:2230td}) show significant dispersions for the
time delays at separate frequencies. If we restrict our consideration to
the approximating functions with the smallest residuals (0528+134 -- D, E;
2230+114 -- A3, B3), the corresponding mean value is
$\alpha_{mean}=(-0.85\pm0.27)$.
Both the ``pessimistic'' and ``optimistic'' estimates for the mean time-lag
index are consistent with the expected value of --0.8, taking into account
the fairly large errors. The only exception might be the index for 1730--130.

\begin{table}
\begin{center}
\vspace{2mm}
\caption{\bf{Parameters of the time-delay function for core outbursts}}
\begin{tabular}{lllll}
\hline
Source &  Comp.  &  $T_{max}$(37GHz) & $\alpha$ \\
0458-020 &  A  &   1989.1  &  -1.46  \\
0528+134 &  D  &   1995.9  &  -0.87  \\
0528+134 &  E  &   1998.4  &  -1.22  \\
0528+134 &  F  &   1999.8  &  -1.31  \\
1730-130 &  A  &   1996.3  &  -2.16  \\
2230+114 &  A2 &   1989.6  &  -0.89  \\
2230+114 &  B2 &   1990.5  &  -1.55  \\
2230+114 &  A3 &   1997.5  &  -0.72  \\
2230+114 &  B3 &   1997.9  &  -0.60  \\
\hline
\end{tabular}
\end{center}
\label{tab:summary}
\end{table}

The shift of the core position between two frequencies is expected to
depend on the luminosity of the source (Lobanov 1998):
$$
\Delta r\propto L_{syn}^{2/3}
$$
As above, we expect a similar dependence for the time lags. The relationship
between the amplitudes of the outbursts in two subsequent activity cycles in
2230+114 and their time lags is in qualitative agreement with the expectation
(Fig.~\ref{fig:2230td}; the outbursts in cycle 3 display both higher
amplitudes and larger time delays). However, the
time lags for individual sub-outbursts within a single outburst do not
display a similar relation with the sub-outburst amplitudes (0528+134).
The time lags for the individual sub-outbursts
may depend on the dynamical evolution of the underlying perturbation,
rather than on their amplitude.

\subsection{Outbursts and structural evolution}

As was already shown by Savolainen et al. (2002), the emergences of
all bright components are correlated with local maxima of the light curves.
A comparison of the 37~GHz light curves with the flux densities of the
cores and new superluminal components as functions of time
(Figs.~\ref{fig:0458}, \ref{fig:0528}, \ref{fig:1730},
\ref{fig:2230vlbi})
augments the correlation. The flux densities of both the cores and new-born
jet components reach their highest values near the maxima of the 37~GHz light
curves, after which they fade.

\begin{figure*}
\centering
\begin{minipage}[c]{\textwidth}
   \centering
   \includegraphics[width=0.45\textwidth]{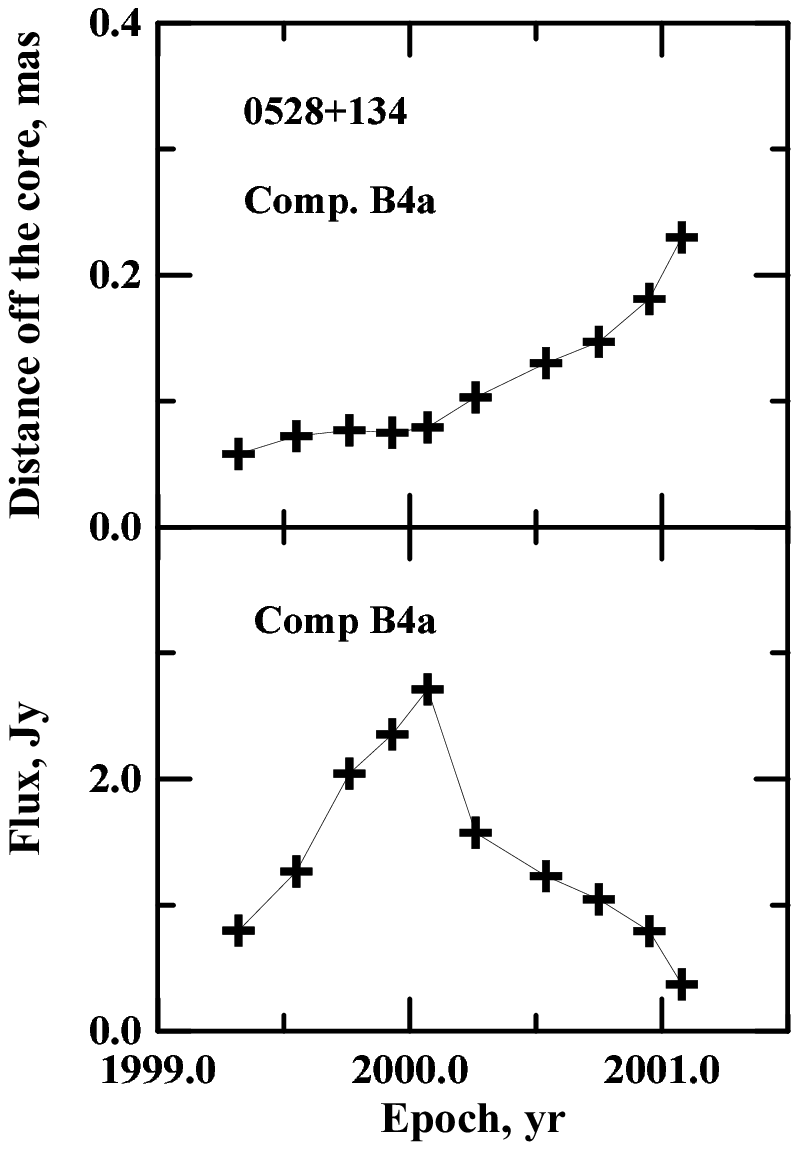}
   \hspace{0.02\textwidth}
   \includegraphics[width=0.45\textwidth]{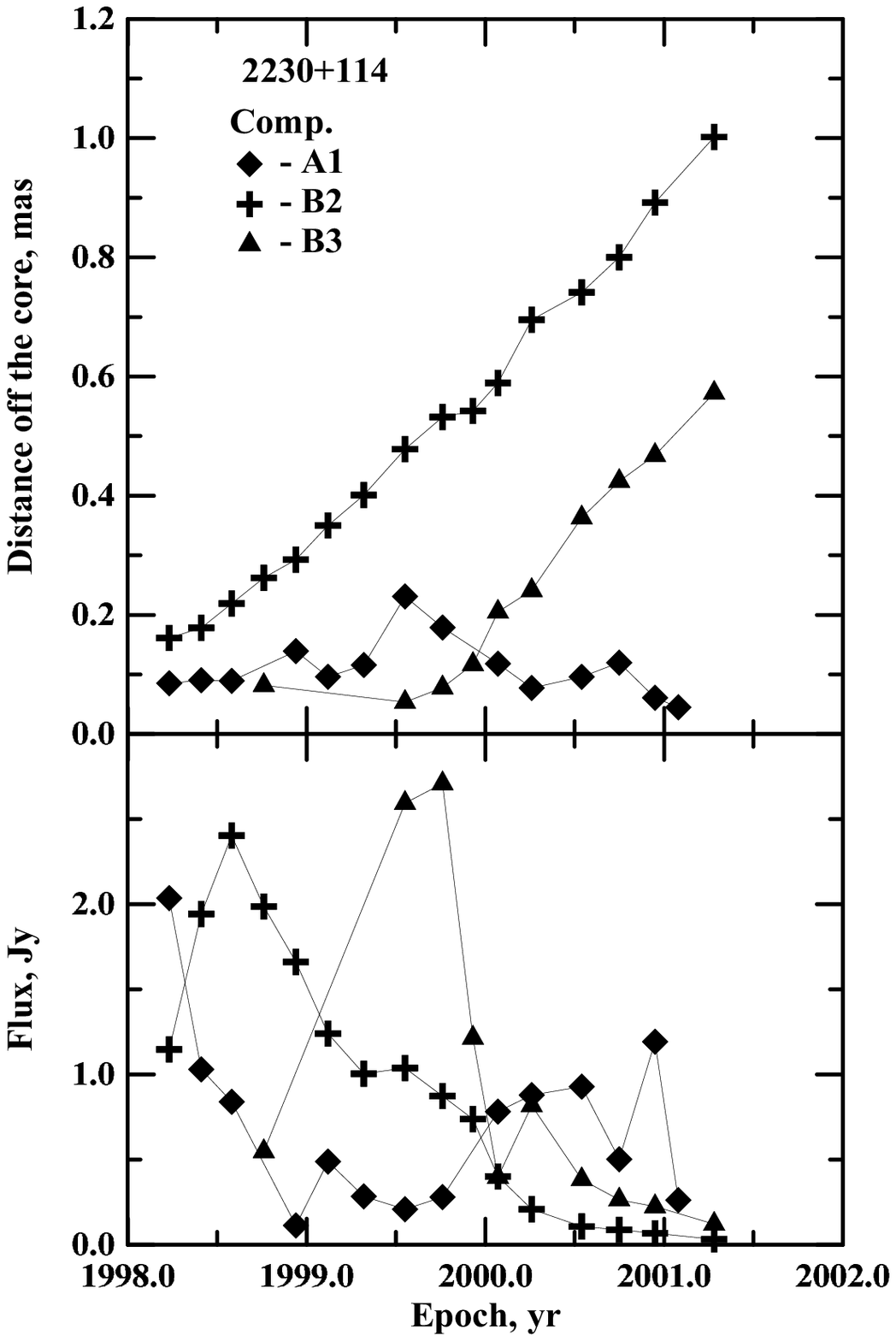}
\end{minipage}
\vspace*{-0.5cm}
\caption{Distance from the core and flux
as functions of time for component B4 in {\bf 0528+134} ({\it left})
and the components A1, B2 and B3 in {\bf 2230+114} ({\it right}) (Jorstad et
al. 2005).}
\label{fig:0528+2230vlbi}
\end{figure*}

Some well studied events (Fig.\ref{fig:0528+2230vlbi}) indicate that the
times when the flux
densities of new-born components reach their maxima nearly coincide with the
times when they experience a sudden increase in their velocity of separation
from the core.
A similar increase in the velocity of the superluminal component C8$^*$
was observed
near 1998 in the quasar 3C279 (Jorstad et al. 2005); note
the close coincidence of this epoch with the maximum of the
brightest outburst observed in 3C279 since 1965. This suggests that the flux
density may reach its maximum when a shocked region that gives birth to a
new jet component becomes optically thin and enters a free-expansion regime,
as has been suggested by Gomez et al. (1997) (although changes
in brightness occuring simultaneously with accelerations can also result
from geometric effects, such as bending of the jet toward or away from the
line of sight; see, e.g., Homan et al. 2003).
Another intriguing phenomenon observed in 2230+114 is the interaction of
the stationary component A1 with the moving component B3, when
A1 is dragged downstream by B3, then returns upstream,
increasing its flux density (Gomez et al. 2001).

\subsection{Conclusions}

Our main results can be summarized as follows.


(1) Frequency-dependent time delays for bright outbursts in four $\gamma$-ray
quasars have been estimated and approximated using an exponential function of
frequency, $T_{max}(\nu)\propto\nu^{\alpha}$. The time delays
in the sources range from $\sim0.3$~yr to $\sim1.0$~yr. The mean value of the
index of the time-delay function is $\alpha_{mean}=(-1.2\pm0.5)$, and the
maximum value in terms of its magnitude is $\alpha=(-2.2\pm0.3)$, for the
source 1730--130.

(2) The observed outbursts in 0528+134 and 2230+114 reveal fine structure and
consist of narrow ($\sim1$ year) sub-outbursts. Some traces of unresolved fine
structure are also seen in 0458--020 and 1730--130. The time delays and
exponential-function indices vary from one sub-outburst to another.

(3) Due to the prominent fine structure observed in some outbursts,
identification of the starting point of the activity requires information
about the spectral evolution at frequencies higher than 40~GHz or the structural
evolution at mm wavelengths.

(4) Lower limits for the durations of the activity cycles, or activity time
scales, have been estimated for 0458--020 ($T_{act}\ge16$~yr) and 1730--130
($T_{act}\ge9$~yr). The results of VLBI imaging at 8~GHz (Britzen et al. 1999b)
and 43~GHz (Jorstad et al. 2001, 2005) suggest that the approximate duration
of the activity cycle for 0528+134 is $T_{act}\sim14$~yr. Bright outbursts 
in 2230+114 repeat with a quasi-period of
$8.04\pm0.30$~yr. Although the amplitudes of individual sub-outbursts vary from
one quasi-period to another, their positions within the overall outburst
are preserved. The time lags for two adjacent periods of activity in 2230+114
suggest a correlation with outburst amplitude: the brighter outburst shows
a greater time delay. The
first maximum of the following bright outburst in 2230+114 is
expected to be near $2005.5\pm0.3$.

(5) The maximum flux densities of new jet components often coincide
with a sudden increase in their velocities from the core. This suggests that
the component's flux reaches its maximum when the shocked region that
corresponds to the jet component becomes optically thin and begins to freely
expand, as was suggested by Gomez et al. (1997).


\section{Acknowledgements}

The work was supported by the RFBR (grant 05-02-17562).
UMRAO has been supported by a series of grants from the NSF and by funds
from the University of Michigan. SGJ was partly supported by the National
Science Foundation grant (AST 04-06865).
This research has made use of the United States Naval Observatory (USNO)
Radio Reference Frame Image Database (RRFID). We acknowledge also the MOJAVE
and VLBA 2-cm Survey Program teams. Finally, we wish to thank the referee,
Matt Lister, for his thorough reading of the manuscript and his helpful
and productive comments for improving this paper.

\section{References}
\noindent
Agudo \i., G\'omez J.L., Gabuzda D.C., Alberdi A., Marscher A. P.,
Jorstad S. G. 2002, Proc. 6th European VLBI Network Symposium,
ed. Ros, E., Porcas, R.W., Lobanov, A.P. \& Zensus, J.A., p. 115

\noindent
Aller H.D., Aller M.F., Latimer G.E. \& Hodge P.E.
1985, ApJS, 59, 513

\noindent
Aller M.
1999, in BL Lac Phenomenon, ed. L.O. Takalo \& A. Sillanp\"a\"a
(San Francisco: ASP), ASP Conf. Ser., 159, p. 31

\noindent
Bower G.C., Backer D.C., Wright M., Forster J.R., Aller H.D. \& Aller M.F.
1997, ApJ, 484, 118

\noindent
Britzen S., Vermeulen R.C., Taylor G.B., Readhead A.C.S., Pearson T.J.,
Henstock D.R. \& Wilkinson P.N.  1999a, in BL Lac Phenomenon,  ed. L.O. Takalo,
A. Sillanp\"a\"a (San Francisco: ASP), ASP Conf. Ser., 159, p. 431

\noindent
Britzen S., Witzel A., Krichbaum T.P., Qian S.J. \& Campbell R.M.
1999b, A\&A, 341, 418

\noindent
Carswell R.F., Strittmatter P.A., Williams R.D., Kinman T.D. \& Serkowski K.
1974, ApJ, 190, L101

\noindent
Dent W.A. \& Kapitzky J.E.
1976, AJ, 81, 1053

\noindent
Fey A.L., Boboltz D.A., Charlot P., Fomalont E.B., Lanyi G.E. \& Zhang L. D.
2005, in Future Directions in High Resolution Astronomy: The 10th Anniversary
of the VLBA, ed. J.D. Romney \& M.J. Reid (San Francisco:ASP), ASP Conf. Ser.,
340, p. 514

\noindent
G\'omez J.L., Alberdi A. \& Marcaide J.M.
1993, A\&A, 274, 55

\noindent
G\'omez J.L., Mart\'i J.M., Marscher A.P., Ib\'a\~nez J.M. \& Alberdi A.
1997, ApJ, 482, L33

\noindent
G\'omez J.L. 2005, in Future Directions in High Resolution Astronomy: The
10th Anniversary of the VLBA, ed. J.D. Romney \& M.J. Reid (San
Francisco:ASP), ASP Conf. Ser., 340, p. 13

\noindent
Hardee P.E., Hughes P.A., Rosen A. \& G\'omez E.A.
2001, ApJ, 555, 744

\noindent
Hewitt A. \& Burbidge G. 1993, ApJS, 87, 451

\noindent
Homan D.C., Lister M.L., Kellermann K.L., Cohen M.H., Ros E., Zensus J.A.,
Kadler M. \& Vermeulen R.C. 2003, ApJ, 589, L9

\noindent
Jorstad S.G., Marscher A.P., Mattox J.R., Aller M.F., Aller H.D.,
Wehrle A.E. \& Bloom S.D. 2001, ApJ, 556, 738

\noindent
Jorstad S.G., Marscher A.P., Lister M.L., Stirling A.M., Cawthorne T.V.,
Gear W.K., Gomez J.L., Stevens J.A., Smith P.S., Forster J.R. \& Robson E.I.
2005, AJ, 130, 1418

\noindent
Kellermann K.I., Vermeulen R.C., Zensus J. A. \& Cohen M.H.
1998, AJ, 115, 1295

\noindent
Kidger M.R., 2000, AJ, 119, 2053

\noindent
Kranich D. 2003, in High Energy Blazar Astronomy, ed. L.O. Takalo \&
E. Valtaoja (San Francisco: ASP), ASP Conf. Ser., 299, p. 3

\noindent
Krichbaum T.P., Britzen S., Standke K.J., Witzel A., Schalinski C.J. \&
Zensus J.A. 1995, Proc. Nat. Acad. Sci. USA, 92 (25), 11377

\noindent
Krichbaum T.P., Witzel A., Graham D. \& Lobanov A.P.
1997, in Millimeter-VLBI Science Workshop, Workshop held 22-23 November
1996 at MIT, ed. R. Barvanis, R.B. Phillips, p. 3

\noindent
Lehto H.J. \& Valtonen M.J.
1996, ApJ, 460, 207

\noindent
Lister M. \& Homan D. 2005, AJ, 130, 1389

\noindent
Lobanov A.P. 1998, A\&A, 330, 79

\noindent
Lobanov A.P., Krichbaum T.P. \& Graham D.A.
2000, A\&A, 364, 391

\noindent
Lobanov A.P. \& Roland J.
2002, Proc. of the 6th European VLBI Network Symposium,
ed. E. Ros, R.W. Porcas, A.P. Lobanov \& J.A. Zensus,
p. 121

\noindent
Lobanov A.P., Hardee P.E. \& Eilek J. 2003
NewAR, 47, 629

\noindent
Marscher A.P. 1996, in Blazar Variability, ed.
H.R. Miller, J.R. Webb \& J.C. Noble (San Francisco, ASP), ASP Conf. Ser.,
110, p. 248

\noindent
Marscher A.P. 2001, in Probing the Physics of Active
Galactic Nuclei by Multiwavelengths Monitoring, ed. B.M. Peterson,
R.S. Polidan, R.W. Pogg (San Francisco, ASP), ASP Conf. Ser., 224, p. 23

\noindent
Marscher A.P., Jorstad S.G., Mattox J.R. \& Wehrle A.E.
2002, ApJ, 577, 85

\noindent
Mukherjee R. et al.  1997, ApJ, 490, 116

\noindent
Mukherjee R., B\"ottcher M., Hartman R.C., Sreekumar P., Thompson D. J.,
Mahoney W. A., Pursimo T., Sillanpää A. \& Takalo L. O.  1999, ApJ, 527, 132

\noindent
Pian E.
2003, in High Energy Blazar Astronomy, ed. L.O. Takalo \& E. Valtaoja
(San Francisco, ASP), ASP Conf. Ser., 299, p. 37

\noindent
Pursimo T. et al.  2000, A\&AS, 146, 141

\noindent
Pyatunina T.B., Marchenko S.G., Marscher A.P., Aller M.F., Aller H.D.,
Ter\"asranta H. \& Valtaoja E.  2000, A\&A, 358, 451

\noindent
Pyatunina T.B., Rachimov I.A., Zborovskii A.A., Gabuzda D.C., Jorstad S.G.,
Ter\"asranta H., Aller M.F. \& Aller H.D.  2003, High Energy Blazar Astronomy,
ed. L.O. Takalo \& E. Valtaoja, (San Francisco:ASP), ASP Conf. Ser., 299, p. 89

\noindent
Raiteri C.M.  et al. 2001, A\&A, 377, 396

\noindent
Reuter H.-P., Kramer C., Sievers A., Paubert G., Moreno R., Greve A.,
Leon S., Panis J.F., Ruiz-Moreno M., Ungerechts H. \& Wild W.
1997, A\&AS, 122, 271.

\noindent
Savolainen T., Wiik K., Valtaoja E., Jorstad S.G. \& Marscher A.P.
2002, A\&A, 394, 851

\noindent
Sillanp\"a\"a A.,
1999, in Observational Evidence for Black Holes in the Universe,
ed. Chakrabarti (Dordrecht: Kluwer), 209

\noindent
Steppe H., Salter, C.J., Chini, R., Kreysa E., Brunswig W. \& Lobato P\'erez J.
1988, A\&AS, 75, 317.

\noindent
Steppe H., Liechti S., Mauersberger R., K\"ompe C., Brunswig W. \& Ruiz-Moreno
M., 1992, A\&AS, 96, 441.

\noindent
Steppe H., Paubert G., Sievers A., Reuter H.P., Greve A., Liechti S.,
Le Floch B., Brunswig W., Men\'endez C. \& Sanches S., 1993, A\&AS, 102, 611.

\noindent
Ter\"asranta H. et al.  1998, A\&AS, 132, 305

\noindent
Ter\"asranta H. et al.  2004, A\&A, 427, 769

\noindent
Ter\"asranta H., Wiren S. \& Koivisto P.  2003, in High Energy Blazar
Astronomy, ed. L.O. Takalo \& E. Valtaoja (San Francisco, ASP), ASP
Conf. Ser., 299, p. 235

\noindent
Ter\"asranta H., Wiren S., Koivisto P., Saarinen V. \& Hovatta T. 2005,
A\&A, 440, 409

\noindent
Valtaoja E., L\"ahteenm\"aki A, Ter\"asranta H. \& Lainela M. 1999,
ApJS, 120, 95

\noindent
Valtaoja E., Ter\"asranta H., Tornikoski M., Sillanp\"a\"a A., Aller M.F.,
Aller H.D. \& Hughes P.A.  2000, ApJ, 531, 744

\noindent
Wright A.E., 1984, Proc. Austr. Astron. Soc., 5, 510

\noindent
Zhang Y.F., Marscher A.P., Aller H.D., Aller M.F., Ter\"asranta H. \&
Valtaoja E.  1994, ApJ, 432, 91

\noindent
Zhou J.F., Hong X.Y., Jiang D.R. \& Venturi T.
2000, ApJ, 540, L13

\end{document}